\documentclass[12pt]{article}
\usepackage{fullpage} 
\usepackage{authblk} 
\usepackage{xcolor}
\providecommand{\keywords}[1]
{
  \small	
  \textbf{\textit{Keywords---}} #1
}

\usepackage{amsmath,amssymb, amsfonts}
\usepackage{comment}
\usepackage{numprint}
\usepackage{float}
\usepackage{natbib}
\usepackage[title]{appendix}
\usepackage{graphicx} 
\usepackage{subcaption}
\graphicspath{ {./Fig/} }
\usepackage{titletoc}
\usepackage{booktabs}
\usepackage[normalem]{ulem}

\DeclareMathOperator*{\V}{\mathbb{V}}

\usepackage{algorithm}
\usepackage{algpseudocode}

\title{Penalized Copula Mixed Models for Intercompany Loss Reserving and Risk Capital}
 \author[1]{Pengfei Cai}
\author[1,2]{Anas Abdallah}
 \author[1,2]{Pratheepa Jeganathan}
 \affil[1]{Department of Mathematics \& Statistics, McMaster University}
 \affil[2]{School of Computational Science and Engineering, McMaster University}
\date{}

\begin{document}

\maketitle

\begin{abstract}
Intercompany loss reserving provides an opportunity to improve reserve estimation by borrowing information across insurers while accounting for company-level heterogeneity. We propose a penalized generalized copula mixed model for multivariate loss reserving and risk capital analysis using multiple companies' loss triangles. The framework combines mixed-effects marginal models with company-specific copula dependence parameters, allowing residual dependence between lines of business to vary across insurers. Penalization is introduced through an $L_1$ penalty on the fixed effects to stabilize estimation in the tail of the loss triangles, where observations are limited.

Estimation is carried out using an iterative two-stage procedure that combines likelihood-based estimation of the marginal mixed models with rank-based copula estimation using residual pseudo-observations. To obtain predictive reserve distributions, we develop a modified bootstrap procedure that accommodates penalized estimation while preserving the dependence structure. Using Schedule P data from the National Association of Insurance Commissioners, we show that the proposed model provides a more stable decomposition of unpaid losses across lines of business, reduces predictive variability relative to silo and fixed-effect copula benchmarks, and leads to lower risk capital after accounting for diversification. A simulation study further evaluates parameter recovery, sparsity selection, reserve accuracy, and robustness to random-effect misspecification. Overall, the proposed model offers an interpretable and flexible framework for intercompany multivariate reserving and capital assessment.
\end{abstract}

\keywords{Loss reserving; Copula mixed models; Intercompany reserving; Risk capital; Penalized estimation}

\section{Introduction}

Intercompany loss triangle analysis is increasingly important for benchmarking reserving practices, supporting regulatory perspectives, and informing market-wide risk management. In such settings, the data consist of triangles observed across multiple insurers, creating a natural opportunity to learn systematic development patterns while accounting for insurer heterogeneity. At the same time, dependence modeling between lines of business (LoBs) remains critical for obtaining predictive reserve distributions and downstream tail metrics used in risk capital and diversification assessment. Our primary statistical object is the predictive distribution of unpaid losses; risk-capital measures (e.g., TVaR) and diversification gains are reported as direct functionals of this  distribution.

This paper is framed as a joint reserving--capital framework for intercompany analysis. Intercompany perspectives have received relatively limited attention in the loss reserving literature, with notable examples including \cite{antonio2010multilevel} and \cite{shi2017multivariate}. Nevertheless, multivariate reserving models that simultaneously capture cross-insurer heterogeneity, dependence across lines of business, and sparsity-inducing regularization remain limited.

\cite{Kuo2019} introduced the Deep Triangle (DT), a recurrent neural network framework that leverages loss triangles from multiple companies to improve reserve prediction. Building on this, the Extended Deep Triangle (EDT) incorporates dependence between two LoBs by modeling pairwise and sequential relationships in loss ratios \citep{cai2025edt}. To account for insurer heterogeneity, EDT encodes company identifiers and applies dropout regularization to induce sparsity. Trained on multiple-company data, EDT produces reserve estimates that closely align with actual reserves while learning development patterns and latent dependence structures directly from the data.

Incorporating dependence between LoBs can improve diversification assessment and reduce estimated risk capital. For example, \cite{cai2025edt} show that modeling pairwise and sequential dependence in multiple loss triangles leads to lower risk capital estimates. In contrast, \cite{Abdallah2015} demonstrate that parametric dependence models provide greater interpretability of dependence structures and diversification effects, although they may sacrifice some predictive accuracy.

Neural network-based reserving methods can generate predictive reserve distributions using generative adversarial network (GAN) techniques, whereas parametric reserving models typically rely on model-based, residual-based, or rank-based bootstrap procedures. In the presence of dependence, copula GANs have been shown to achieve substantial risk-capital gains \citep{cai2025edt}, while rank-based bootstrapping of copula coordinates can provide meaningful diversification benefits among parametric reserving approaches \citep{abdallah2023rank}.

Parametric dependence models provide interpretable dependence parameters and facilitate reserve and capital estimation. Copula regression offers a flexible framework for modeling dependence between LoBs, but most applications focus on single-company settings or treat company effects as fixed. For example, \cite{zhang2010general} proposes a multivariate seemingly unrelated regression framework, while \cite{Shi2011} and \cite{Abdallah2015} employ copula regression with exponential-family marginals, which is important for the heavy-tailed nature of insurance losses.

A natural extension of copula reserving models is to pool triangles from multiple insurers. However, the use of a common dependence parameter implicitly assumes homogeneous diversification effects across insurers. Allowing dependence to vary across companies recognizes that diversification effects may differ substantially between insurers due to differences in portfolio composition, underwriting practices, and claims management. To address intercompany reserving, \cite{shi2017multivariate} proposed a Bayesian hierarchical copula model that borrows strength across insurers. Nevertheless, the marginal models are restricted to a log-normal specification, and insurer-specific dependence structures are not explicitly modeled.

Applications of random effects in multivariate loss reserving remain relatively limited \citep{Abdallah2016}. More recently, \cite{karatacs2025dependence} incorporated random effects within a copula regression framework to model unobserved variation between and within lines of business. However, applications combining mixed effects, copula dependence, and insurer-specific diversification remain limited.

Regularization is particularly useful in aggregate loss reserving because the number of observed cells decreases in the most recent accident and development years, where reserve extrapolation is most important. In such settings, accident-year and development-year effects may be weakly identified or estimated from limited data. Shrinkage methods such as the least absolute shrinkage and selection operator (LASSO) \citep{tibshirani1996regression} can stabilize these effects and yield more parsimonious reserving models. In this paper, sparsity refers to coefficient sparsity induced by the $L_1$ penalty.

\cite{mcguire2018self} introduced a LASSO framework for loss reserving using individual claim data, applying $L_1$ regularization to stabilize estimation. Regularization is particularly beneficial in reserving applications where collinearity and limited observations can compromise model stability. Moreover, LASSO can be readily incorporated into copula regression, enabling variable selection while preserving predictive performance.

In this paper, interpretability is understood relative to black-box multivariate architectures: the proposed structure decomposes reserve dynamics into (i) systematic accident-year and development-year effects, (ii) insurer-level heterogeneity summarized through random effects, and (iii) a copula dependence parameter that directly quantifies cross-LoB association after conditioning on (i)–(ii). Sparsity induced by the LASSO further improves interpretability by removing negligible calendar effects, yielding a parsimonious specification that is easier to communicate and stress-test in practice.

Motivated by these limitations, we propose a penalized generalized copula mixed model (pGCMM) for intercompany loss triangles that extends copula regression to accommodate hierarchical data from multiple companies. The model combines fixed effects for accident-year and development-year patterns, company-specific random effects for insurer heterogeneity, and a copula component for residual cross-LoB dependence. Sparsity is introduced through $L_1$ regularization on the fixed effects, improving model parsimony and stability. This decomposition separates insurer-level heterogeneity from residual cross-LoB dependence, providing an interpretable framework that borrows strength across companies while retaining company-specific diversification effects. This formulation can be viewed as a credibility-type extension of copula regression, where information is partially shared across companies through the random-effect distribution.

Maximum likelihood estimation of copula mixed models is computationally challenging because the copula term is embedded within the random-effects likelihood. Moreover, copula estimation may be sensitive to marginal misspecification. We therefore adopt a two-stage estimation strategy that combines likelihood-based estimation of the marginal mixed models with rank-based pseudo-observations for copula estimation. The resulting dependence parameter measures residual cross-LoB association after removing systematic development patterns and insurer-level heterogeneity.

Bootstrap methods are widely used to generate predictive reserve distributions from incremental paid losses \citep{davison1997bootstrap,kirschner2008two,taylor2007synchronous}. However, standard bootstrap procedures may be unstable in penalized models when some fixed-effect coefficients are shrunk to zero. To address this issue, we adopt the modified bootstrap of \cite{chatterjee2011bootstrapping} and combine it with residual-rank resampling to accommodate sparsity while preserving the dependence structure. The resulting procedure provides a stable estimate of the predictive reserve distribution under the proposed penalized copula mixed model.

Although our empirical illustration focuses on two LoBs, the framework extends naturally to more than two lines. In this setting, the pGCMM becomes a system of mixed-effects regressions, and the dependence structure can be modeled using higher-dimensional elliptical or vine copulas. Estimation may follow the same two-stage strategy based on marginal mixed models and residual-rank dependence estimation.

We summarize our contributions as follows

\begin{enumerate}
    \item We introduce a generalized copula mixed modeling (GCMM) framework for intercompany multivariate reserving, enabling estimation of predictive reserve distributions and associated capital functionals under cross-insurer heterogeneity and cross-LoB dependence.
    \item We allow the copula dependence parameters to vary by company, thereby capturing insurer-specific diversification patterns that may be obscured under a common dependence parameter.
    \item We propose an iterative two-stage estimation procedure that combines marginal mixed-model estimation with rank-based copula estimation using residual pseudo-observations.
    \item We incorporate an $L_1$ penalty on accident-year and development-year effects to stabilize weakly identified fixed effects in the tail of the loss triangles and obtain a more parsimonious reserving model.
    \item We adapt residual bootstrapping to generate predictive reserve distributions and evaluate reserve uncertainty, TVaR, and risk capital.
\end{enumerate}

The paper is organized as follows: Section 2 introduces the GCMM and its penalized extension (pGCMM), along with the estimation procedure and the modified bootstrap approach for generating predictive reserve distributions. Section 3 applies and calibrates the proposed pGCMM using a dataset that includes personal and commercial automobile LoBs from multiple companies. Section 4 presents a simulation study evaluating parameter recovery, reserve accuracy, and robustness under controlled data-generating settings. Finally, Section 5 summarizes our main findings and concludes.

\section{Generalized Copula Mixed Model}

\subsection{Model specification}

Let $Y_{ijc}^{(\ell)}$ denote the standardized incremental paid losses for line of business (LoB) $\ell$,  $\ell \in \lbrace 1, 2, \ldots, L \rbrace$, accident year $i$ and development year $j$, $i, j \in \lbrace 1, 2, \ldots, I \rbrace$, and company $c$, $c \in \lbrace 1, 2, \ldots, C \rbrace$. In this paper, we specify a generalized copula mixed model (GCMM) for two LoBs with company-specific dependence parameters. 

We assume that conditional on the company-specific random effect $b_{c}^{(\ell)}$, the standardized incremental paid losses $Y_{ijc}^{(\ell)}$ follow a two-parameter distribution from the exponential family (e.g., gamma or log-normal) as

$$Y_{ijc}^{(\ell)} \mid b_{c}^{(\ell)} \sim f\left(y_{ijc}^{(\ell)} \mid \mu_{ijc}^{(\ell)}, \sigma_{\ell}\right),$$
allowing both the mean and dispersion (variance or shape) to be modeled.
For the applications, we choose the marginal distribution for each LoB using the Akaike Information Criterion (AIC) or Bayesian Information Criterion (BIC). 

The conditional mean $\mu_{ijc}^{(\ell)}$ is defined via the LoB-specific link function $g_{\ell}(\cdot)$ as

\begin{equation}    g_{\ell}(\mu_{ijc}^{(\ell)}) = \eta_{ijc}^{(\ell)}=\boldsymbol{x}_{ij}^{(\ell)}\boldsymbol{\beta}^{(\ell)}+ b^{(\ell)}_{c},
\end{equation}
where $\boldsymbol{x}_{ij}^{(\ell)}$ is a $(2I-1)$-dimensional row vector of the design matrix to a specific cell in the loss triangle. For example, $\boldsymbol{x}_{12}^{(\ell)}
= \left(1,\underbrace{ 0, \ldots, 0}_{I-1},\underbrace{1, 0, \ldots, 0}_{I-1} \right)^{\top}$. It consists of indicator variables that map the specific observation $y_{ijc}^{(\ell)}$ at accident year $i$ and development year $j$ to the corresponding effects in $\boldsymbol{\beta}^{(\ell)}$. 

$\boldsymbol{\beta}^{(\ell)}$ contains all the systematic component parameters that apply to every company triangle, particularly $\boldsymbol{\beta}^{(\ell)} = \left(\xi^{(\ell)}, \alpha^{(\ell)}_{2}, \ldots, \alpha^{(\ell)}_{I}, \beta^{(\ell)}_{2}, \ldots, \beta^{(\ell)}_{I}\right)^{\top}$, $\alpha_{1}^{(\ell)}=0$ and $\beta_{1}^{(\ell)}=0$. We assume that $\boldsymbol{\beta}^{(\ell)}$ captures systematic trends in accident and payment patterns in LoB $\ell$ industry-wide. 

$b^{(\ell)}_{c}$ is a company-specific intercept for company $c$ in LoB $\ell$. $b^{(\ell)}_{c}$ is a realization from the distribution $\text{N}\left(0, \tau^{2}_{\ell}\right)$, where $\tau^{2}_{\ell}$ is the variance of company random effect in LoB $\ell$. 

To capture the dependence between LoBs, we apply a copula to the conditional marginals. $b^{(\ell)}_{c}$ handles whether a specific company may consistently over- or under-reserve compared to all companies on average.

Based on Sklar's theorem \citep{Nelsen_copula_2006}, for a specified cell $\left(i, j, c\right)$, the joint density of the vector $\boldsymbol{Y}_{ijc} = \left(Y_{ijc}^{(1)}, Y_{ijc}^{(2)}\right)^{\top}$ given the vector of random effects $\boldsymbol{b}_{c} = \left(b_{c}^{(1)}, b_{c}^{(2)}\right)^{\top}$ is 

\begin{equation}
\label{eq:yijc_con-cop}
    f\left(\boldsymbol{y}_{ijc} \mid\boldsymbol{b}_{c}\right) = \left[\prod_{l=1}^{2}f_{\ell}(y_{ijc}^{(\ell)} \mid b_{c}^{(\ell)}, \boldsymbol{\beta}^{(\ell)}, \sigma_{\ell}) \right] c\left(u_{ijc}^{(1)}, u_{ijc}^{(2)}; \theta_{c}\right),
\end{equation}
where $u_{ijc}^{(\ell)}$ is a realization of $U_{ijc}^{(\ell)} = F_{\ell}\left(y_{ijc}^{(\ell)} \mid b_{c}^{(\ell)},  \boldsymbol{\beta}^{(\ell)}, \sigma_{\ell}\right)$, and $F_{\ell}$ denotes the conditional cumulative distribution function (CDF) of LoB $\ell$. $c\left(\cdot ; \theta_{c}\right)$ is a copula density that models the seemingly unrelated regression-type dependence, but using probability integral transformation of $\boldsymbol{Y}_{ijc}  \mid \boldsymbol{b}_{c},  \boldsymbol{\beta}^{(\ell)}, \sigma_{\ell}$ and company-specific parameter $\theta_c$. This captures residual stochastic variation that affects multiple LoBs simultaneously and remains after accounting for accident-year effects, development-year effects, and company-specific heterogeneity, not the dependence due to common development patterns or company-specific effects.

\subsection{ Likelihood and Rank-based Pseudo-Likelihood}

The likelihood is constructed by integrating the joint conditional likelihood over the company-specific random effects. For two LoBs, the random-effect vector is $\left(b_{c}^{(1)}, b_{c}^{(2)}\right)^{\top}$, and the integration is therefore performed over $\mathbb{R}^2$. Conditional on $\boldsymbol{b}_{c}$, $\boldsymbol{Y}_{ijc} = \left(Y_{ijc}^{(1)}, Y_{ijc}^{(2)}\right)^{\top}$ are assumed independent across triangle cells $(i,j)$ within a company, while dependence between LoBs at a given cell is captured by the copula. If $\boldsymbol{b}_{c} = \left(b_{c}^{(1)}, b_{c}^{(2)}\right)^{\top} \sim \text{N}\left(\boldsymbol{0}, \tau\right)$, $\tau = \text{diag}\left(\tau^{2}_{1}, \tau^{2}_{2}\right)$, the log-likelihood $\mathcal{L}(\boldsymbol{\Theta})$ is
\begin{equation}
\notag
 \mathcal{L}(\boldsymbol{\Theta}) = \sum_{c=1}^{C}\log \int_{\mathbb{R}^2} \left( \prod_{i=1}^{I} \prod_{j=1}^{I+1-i} f\left(\boldsymbol{y}_{ijc} \mid \boldsymbol{b}_{c}, \boldsymbol{\beta}, \boldsymbol{\sigma}, \theta_{c}\right) \right) f\left(\boldsymbol{b}_{c}; \tau \right) d \boldsymbol{b}_{c}. 
\end{equation}
Expanding this with the copula term in Eq. \eqref{eq:yijc_con-cop},
\begin{equation}
    \label{eq:full-log-lik}
    \mathcal{L}(\boldsymbol{\Theta}) = \sum_{c=1}^{C}\log \int_{\mathbb{R}^2} \left( \prod_{i=1}^{I} \prod_{j=1}^{I+1-i}  \left[\prod_{l=1}^{2}f_{\ell}(y_{ijc}^{(\ell)} \mid b_{c}^{(\ell)}, \boldsymbol{\beta}^{(\ell)}, \sigma_{\ell}) \right] c\left(u_{ijc}^{(1)}, u_{ijc}^{(2)}; \theta_{c}\right)\right) f\left(\boldsymbol{b}_{c}; \tau \right) d \boldsymbol{b}_{c}, 
\end{equation}
where $\boldsymbol{\Theta} = \lbrace\boldsymbol{\beta}^{(1)},  \boldsymbol{\beta}^{(2)}, \sigma_{1}, \sigma_{2},\tau_{1}, \tau_{2}, \theta_{1}, \ldots, \theta_{C}\rbrace$.

The log-likelihood in Eq.~\eqref{eq:full-log-lik} defines the hierarchical copula mixed model. Direct maximization of this log-likelihood is computationally demanding because the copula term is embedded inside the integral over the company-specific random effects. 

\color{black}

\subsection{Estimation Procedure}\label{esti-proc}

Next, we describe an iterative two-step approach to estimate the model parameters using the likelihood formulation in Eq.~\eqref{eq:full-log-lik} together with a rank-based pseudo-likelihood for the copula step.

We use the log-likelihood as the model-defining object, but estimate the parameters through an iterative two-stage procedure. The marginal mixed-model parameters are estimated conditional on the current dependence parameter, and the copula parameter is updated using rank-based pseudo-observations constructed from residuals after removing systematic and company-level effects. This approach preserves the interpretation of the copula as residual cross-LoB dependence while avoiding spurious dependence induced by fixed effects or company-specific shifts.

In Step 1, we estimate the marginal model parameters $\lbrace \boldsymbol{\beta}^{(1)}, \boldsymbol{\beta}^{(2)}, \sigma_{1}, \sigma_{2}, \tau_{1}, \tau_{2}\rbrace$ while holding the company-specific copula parameters fixed. The integrated log-likelihood is approximated using $M$ Monte Carlo samples from the random-effect distributions $b_{c,m}^{(\ell)} \sim \text{N}\left(0, \tau_{\ell}^{2}\right)$:
\begin{equation}
    \label{eq:app-step1}
    \mathcal{L}(\boldsymbol{\Theta}_{-\theta_c}) \approx \sum_{c=1}^{C}\log \left[ \dfrac{1}{M} \sum_{m=1}^{M}\left(  \prod_{i=1}^{I} \prod_{j=1}^{I+1-i} \left[\prod_{l=1}^{2}f_{\ell}(Y_{ijc}^{(\ell)} \mid b_{c,m}^{(\ell)}, \boldsymbol{\beta}^{(\ell)}, \sigma_{\ell}) \right] c\left(u_{ijc}^{(1)}, u_{ijc}^{(2)}; \theta_{c}\right)\right) \right],
\end{equation}
where $\boldsymbol{\Theta}_{-\theta_c}$ denotes a collection of all model parameters except the company-specific copula parameter $\theta_c$. In the first iteration, the copula parameters are initialized at $\theta_c = 0$, corresponding to independence between the two LoBs. Eq. \eqref{eq:app-step1} is then optimized with respect to $\boldsymbol{\Theta}_{-\theta_c}$.  In subsequent iterations, the current estimates of $\theta_c$ obtained from Step 2 are treated as fixed, and the marginal model parameters are updated. At each outer iteration, the pseudo-observations (defined below) are recomputed using the updated marginal parameter estimates before proceeding to Step 2.

In Step 2, with the marginal model parameters fixed, we isolate the residual stochastic dependence between the two LoBs. We first compute pseudo-residuals $\epsilon_{ijc}^{(\ell)}$ using the parameter estimates obtained in Step 1. If $Y_{ijc}^{(\ell)} \mid b_{c}^{(\ell)}$ follows a log-normal distribution, the pseudo-residuals \citep[pp.~331-340]{davison1997bootstrap} are computed as 
\begin{equation}
\epsilon_{ijc}^{(\ell)}=\dfrac{\log y_{ijc}^{(\ell)}-\hat{\mu}_{ijc}^{(\ell)}}{\hat{\sigma}_{\ell}},
\label{eqn:lognormal_residual}
\end{equation}
where $\hat{\mu}_{ijc}^{(\ell)}$ and $\hat{\sigma}_{\ell}$ are estimates. If $Y_{ijc}^{(\ell)}$ follows a gamma distribution, the pseudo-residuals are computed as
\begin{equation}
\epsilon_{ijc}^{(\ell)}=\dfrac{y_{ijc}^{(\ell)}-\hat{\mu}_{ijc}^{(\ell)}}{\sqrt{\widehat{\V} \left(Y_{ijc}\right)}},
\label{eqn:gamma_residual}
\end{equation}
where $\hat{\mu}_{ijc}^{(\ell)}$ and $\hat{\gamma}_{ijc}^{(\ell)}$ are estimates from Step 1 and $\widehat{\V}\!\left(Y_{ijc}^{(\ell)}\right)=\hat{\mu}_{ijc}^{(\ell)}\hat{\gamma}_{ijc}^{(\ell)}$ under the adopted gamma parametrization.

We then transform the pseudo-residuals into pseudo-observations using their empirical ranks within each company and LoB. The rank $R_{ijc}^{(\ell)}$ of the pseudo-residual $\epsilon_{ijc}^{(\ell)}$ is given by 
\begin{equation}
    R_{i jc}^{(\ell)}=\frac{1}{I(I+1)/2+1} \sum_{i^*=1}^{I} \sum_{j^*=1}^{I+1-i^*} \mathbf{1}\left(\varepsilon_{i^* j^*c}^{(\ell)} \leq \varepsilon_{i jc}^{(\ell)}\right),
    \label{eq:pseudo_res}
\end{equation}
where $\mathbf{1}$ is the indicator function.

The company-specific copula parameters are then updated by replacing the conditional CDF values in the copula density with the rank-based pseudo-observations and maximizing the resulting pseudo-log-likelihood, while holding the marginal model parameters fixed.

\begin{equation}
    \label{eq:appr-Step2}
    \mathcal{L}(\theta_{c}) \approx \sum_{c=1}^{C} \log \dfrac{1}{M} \sum_{m=1}^{M} \left( \prod_{i=1}^{I} \prod_{j=1}^{I+1-i}  \left[\prod_{l=1}^{2}f_{\ell}(Y_{ijc}^{(\ell)} \mid b_{c,m}^{(\ell)}, \boldsymbol{\beta}^{(\ell)}, \sigma_{\ell}) \right] c\left(R_{ijc}^{(1)}, R_{ijc}^{(2)}; \theta_{c}\right)\right), 
\end{equation}
where $R_{ijc}^{(1)}$ and $R_{ijc}^{(2)}$ denote the pseudo-observations obtained from the residual ranks. The updated estimates of $\theta_{c}$ are then passed back to Step 1.

The algorithm iterates between Step 1 and Step 2 until convergence. Let $\boldsymbol{\Theta}_{-\theta_c}^{(r)}$ and
$\boldsymbol{\theta}^{(r)}=(\theta_1^{(r)},\ldots,\theta_C^{(r)})$
denote the parameter estimates at iteration $r$. Iteration is stopped when both
\begin{equation}
    \dfrac{1}{p}\sum_{k=1}^{p}\left|\Theta_{-\theta_c,k}^{(r+1)}-\Theta_{-\theta_c,k}^{(r)}\right| < \varepsilon_1
    \label{eq:conv_pars}
\end{equation}
and
\begin{equation}
\max_{c=1,\ldots,C} \left| \theta_c^{(r+1)}-\theta_c^{(r)}\right| < \varepsilon_{2},
\label{eq:conv_theta}
\end{equation}
where $p$ denotes the number of marginal model parameters and $\varepsilon_1$ and $\varepsilon_{2}$ are pre-specified tolerances.

As starting values for $\boldsymbol{\beta}$, separate marginal models are fitted to each LoB, and the resulting estimates are used to initialize the fixed effects, dispersion parameters, and random-effect standard deviations. The complete estimation algorithm is provided in Algorithm~\ref{alg:gcmm-estimation}.

\subsection{Penalized Generalized Copula Mixed Model }\label{sparse-esti-proc}


In this section, we incorporate the Least Absolute Shrinkage and Selection Operator (LASSO) into the generalized copula mixed model.

We define the penalized loss function, $J_{\lambda_1, \lambda_2}(\boldsymbol{\Theta})$, as the negative log of the integrated joint likelihood in Eq. \eqref{eq:full-log-lik} combined with an $L_1$ penalty on the fixed effects. We deliberately refrain from penalizing the copula parameters $\theta_c$ to avoid biasing the captured residual stochastic dependence toward independence. The penalized loss function is defined as 
\begin{equation}
    \label{eq:pen-neg-loglik}
    J_{\lambda_1, \lambda_2} (\boldsymbol{\Theta}) = -\sum_{c=1}^{C} \log \int_{\mathbb{R}^2} \underbrace{\left( \prod_{i=1}^{I} \prod_{j=1}^{I+1-i} f(\mathbf{y}_{ijc} \mid \mathbf{b}_c, \boldsymbol{\beta}, \boldsymbol{\sigma}, \theta_c) \right)}_{\text{Conditional copula density}} f(\mathbf{b}_c ; \boldsymbol{\tau}) \, d\mathbf{b}_c + \sum_{\ell=1}^2 \lambda_\ell ||\boldsymbol{\beta}^{(\ell)}||_1.
\end{equation}
Expanding this using the Monte Carlo approximation, the objective function becomes
\begin{equation}
  \label{eq:app-pen-neg-loglik}
  J_{\lambda_1, \lambda_2} (\boldsymbol{\Theta}) \approx -\sum_{c=1}^{C} \log \left( \frac{1}{M} \sum_{m=1}^{M} \prod_{i=1}^{I} \prod_{j=1}^{I+1-i} f(\mathbf{y}_{ijc} \mid \mathbf{b}_{c,m}, \boldsymbol{\beta}, \boldsymbol{\sigma}, \theta_c) \right) + \sum_{\ell=1}^2 \lambda_\ell ||\boldsymbol{\beta}^{(\ell)}||_1 \ .
\end{equation}
Estimation follows the iterative two-step procedure described in Section \ref{esti-proc}, with the only modification occurring in Step 1. Specifically, for fixed company-specific copula parameters $\theta_c$, the marginal model parameters are estimated by minimizing the penalized objective function in Eq. \eqref{eq:app-pen-neg-loglik}. The $L_{1}$ penalty induces sparsity in the accident-year and development-year effects while retaining the random-effect and dispersion parameters. Because the penalty is non-differentiable at zero, optimization is performed using coordinate descent. The tuning parameters $\lambda_1$ and $\lambda_2$ control the degree of sparsity for LoBs 1 and 2, respectively.

Step 2 remains unchanged. Given the penalized marginal model parameter estimates obtained from Step 1, pseudo-residuals and pseudo-observations are recomputed and the company-specific copula parameters $\theta_c$ are updated using the rank-based pseudo-likelihood in Eq. \eqref{eq:appr-Step2}. The algorithm alternates between the penalized Step 1 and the copula update in Step 2 until the convergence criteria described in Section \ref{esti-proc} are satisfied.

The optimal regularization parameters $\left(\lambda_1, \lambda_2\right)$ are selected by minimizing the Akaike Information Criterion, which balances the joint model fit with the effective  degrees of freedom, defined as

\begin{equation}
\label{eqn:aic_sparse_model}
    \text{AIC}=-2 \log L (\hat{\boldsymbol{\beta}}^{(1)},\hat{\boldsymbol{\beta}}^{(2)} ,\hat{\sigma}_1,\hat{\sigma}_2, \hat{\tau}_1,\hat{\tau}_2, \hat{\theta_c}) + 2 \cdot \hat{df}_{\lambda_1,\lambda_2},
\end{equation}
where $\hat{df}_{\lambda_1,\lambda_2}=\sum_{\ell=1}^{2}\left|\left\{k :\hat{\beta}^{(\ell)}_k \neq 0 \right\}\right| +2+2+C$, accounting for the number of nonzero fixed effect coefficients, the two variance parameters, the two dispersion parameters, and the dependence parameters. 
The complete estimation algorithm is provided in Algorithm~\ref{alg:pgcmm-estimation}.

\subsection{Predictive Distribution of the Total Reserve}\label{sec:boot_procedure}

We describe a modified bootstrap procedure for the pGCMM following \cite{chatterjee2011bootstrapping}. First we fit the pGCMM to the observed incremental paid losses $y_{ijc}^{(\ell)}$ and obtain the optimal penalization parameters $\hat{\lambda}_1$, $\hat{\lambda}_2$, and model parameters $\hat{\boldsymbol{\beta}}^{(\ell)}$,$\hat{\sigma}_\ell$, $\hat{\tau}_\ell$, and $\hat{\theta}$.

Next, we construct thresholded coefficient estimates $\tilde{\boldsymbol{\beta}}^{(\ell)}$ following the modified bootstrap idea of \citet{chatterjee2011bootstrapping}. Specifically, coefficients with absolute values less than 0.001 are set to zero. The resulting thresholded estimates $\tilde{\boldsymbol{\beta}}^{(\ell)}$ are used only for generating bootstrap samples and computing the associated pseudo-residuals.

Then, using $\tilde{\boldsymbol{\beta}}^{(\ell)}$, $\hat{\sigma}_\ell$, $\hat{\tau}_\ell$, $\hat{\theta}_{c}$, and the posterior mode estimates $\hat{b}_{c}^{(\ell)}$, we compute fitted means $\tilde{\mu}_{ijc}^{(\ell)}=\exp (\boldsymbol{x}_{ij}^{(\ell)}\tilde{\boldsymbol{\beta}}^{(\ell)}+ \hat{b}_{c}^{(\ell)}),$
where $\hat{b}_{c}^{(\ell)}$ denotes the posterior mode of the company-specific random effect for company $c$ and LoB $\ell$ obtained from the fitted pGCMM conditional on the observed data. These fitted means are then used to compute pseudo-residuals as defined in \eqref{eqn:lognormal_residual} or \eqref{eqn:gamma_residual}.

Then, for each company $c$, we simulate pairs $(u_{ijc}^{(1)},u_{ijc}^{(2)})$ from the fitted copula $C(\cdot;\hat{\theta}_{c})$. These pairs are transformed via the empirical quantile function of the pseudo-residuals to obtain $\epsilon_{ijc}^{*(\ell)}$. Finally, bootstrap samples of the incremental paid losses are generated by applying the inverse transformation associated with the marginal-model-specific pseudo-residual definition using $\epsilon_{ijc}^{*(\ell)}$, $\tilde{\mu}_{ijc}^{(\ell)}$, and estimated dispersion parameter $\hat{\sigma}_\ell$. For example, under the gamma marginal model, $y_{ijc}^{*(\ell)}=\epsilon_{ijc}^{*(\ell)}*(\tilde{\mu}_{ijc}^{(\ell)}*\hat{\sigma}_{\ell})^{1/2}+\tilde{\mu}_{ijc}^{(\ell)}$.

Given the bootstrap sample pairs $(y_{ijc}^{*(1)},y_{ijc}^{*(2)})^\top$, we refit the pGCMM described in Section \ref{sparse-esti-proc} using the fixed penalization parameters $\hat{\lambda}_1$ and $\hat{\lambda}_2$. The resulting parameter estimates are used to predict the lower-triangle incremental losses and obtain the reserve estimate for company $c$,
\begin{equation}
    \sum_{\ell=1}^{2} \sum_{i=2}^{I} \sum_{j=I-i+2}^{I} \omega_{ic}^{(\ell)} \cdot \exp  (\hat{\eta}_{ijc}^{*(\ell)}) ,
\end{equation}
where $\hat{\eta}_{ijc}^{*(\ell)}=\boldsymbol{x}_{ij}^{(\ell)}\hat{\boldsymbol{\beta}}^{*(\ell)}+\hat{b}_{c}^{*(\ell)} 
$ and $\omega_{ic}^{(\ell)}$ is the premium for accident year $i$ and company $c$ in LoB $\ell$. We repeat this procedure 5,000 times to obtain an approximation to the predictive distribution of the reserve. The modified bootstrapping procedure is provided in Algorithm~\ref{alg:pgcmm-bootstrap}.
\color{black}

\section{Application}
\label{application}

To evaluate multivariate loss reserving and risk capital estimation in an intercompany setting, we analyze a dataset consisting of 30 pairs of loss triangles obtained from Schedule P of the National Association of Insurance Commissioners (NAIC) database \citep{meyers2011loss}. Each pair corresponds to a single company and contains data for the personal auto and commercial auto lines of business (LoB). Each triangle reports incremental paid losses for accident years 1988 to 1997, spanning ten development years. This dataset is representative of an intercompany setting (triangles observed across insurers) in which benchmarking and market-level inference are of interest, while the model remains usable as a structured “borrowing-strength” approach whenever multiple related portfolios are available.

We first use this dataset to investigate the predictive performance of the proposed penalized generalized copula mixed model (pGCMM) for multivariate loss reserving. We then compare it with the unpenalized generalized copula mixed model (GCMM), a copula regression with company fixed effects (Copula-FE), and a LoB-wise generalized linear mixed model (Silo-GLMM), to assess the effects of predictor selection, insurer heterogeneity, and residual cross-LoB dependence on predictive accuracy and risk capital estimation.

Before comparing the joint reserving models, we first examine the suitability of the marginal distributions for each LoB. For each company and LoB, we fit gamma and log-normal distributions to the observed upper-triangle loss ratios and evaluate goodness-of-fit using Kolmogorov-Smirnov tests. At the 5\% level, the gamma distribution is not rejected for 28 of the 30 companies in personal auto and 19 of the 30 companies in commercial auto, while the log-normal distribution is not rejected for 20 and 9 companies, respectively. These results support using gamma marginals for both LoBs, while also indicating that the log-normal distribution remains plausible for the personal auto LoB. This preliminary step is important because the estimated copula dependence is computed from residual or rank-based pseudo-observations and can therefore be sensitive to marginal misspecification. In the main comparison below, we use gamma marginals for both LoBs to maintain a common marginal specification and thereby attribute differences in predictive performance and risk capital estimates to mixed effects, sparsity, and copula dependence rather than to the choice of marginal distribution.


Although gamma marginals are used throughout the main comparison, alternative marginal choices, especially for personal auto, may affect both reserve levels and the estimated residual dependence. We return to this point in the discussion.


The four models differ in accounting for insurer heterogeneity, residual cross-LoB dependence, and fixed-effect regularization. The first is the proposed pGCMM, in which accident-year and development-year effects are estimated using LASSO regularization. The second is the corresponding unpenalized GCMM. The third, Copula-FE, is a copula regression with company fixed effects. The fourth, Silo-GLMM, is a generalized linear mixed model, fitted separately for each LoB and therefore unable to capture residual cross-LoB dependence through a copula. This comparison allows us to isolate the effects of three modeling choices: accounting for insurer heterogeneity through random effects, modeling residual cross-LoB dependence through a copula, and stabilizing fixed effects through LASSO regularization.

Given 30 pairs of loss triangles, we used a validation approach to assess predictive performance of the models. Five of the 30 companies were randomly selected, and for each selected company the last five observed development lags in the upper triangle were held out for validation. The remaining observed upper-triangle cells were used for model fitting. Using the training set, pGCMM was fitted over a grid of regularization parameters, and the optimal penalty pair was selected by minimizing the AIC. Table \ref{tab:validation} shows that the pGCMM and GCMM exhibited similar predictive performance on the validation set. We therefore use the AIC-selected pGCMM in the subsequent analysis and compare it with GCMM, Copula-FE, and Silo-GLMM. After model selection, each model was refitted to the full dataset and used to obtain bootstrap standard errors, reserve estimates, and risk capital measures. 

For completeness, Appendix Tables \ref{app:one_theta} and \ref{app:one_theta_comparison} report the corresponding common-dependence comparison. This additional comparison illustrates the effect of replacing a single common copula parameter with company-specific dependence parameters, which is central in intercompany reserving applications.

Next, we focus on Company 1, a major U.S. property-casualty insurer, and compare the resulting reserve estimates with the observed lower-triangle payments, which are known because the dataset is historical.

Table~\ref{tab:reserve_point_estimates} reports the point estimates of unpaid losses for personal auto, commercial auto, and the combined portfolio. Table~\ref{tab:reserve_percentage_error} reports the corresponding percentage errors relative to the actual reserve,
\begin{equation}
\text{Percentage Error}_{\ell}
=100\times\frac{\hat{R}^{\ell}-R^{\ell}}{R^{\ell}},
\end{equation}
where $\hat{R}^{\ell}$ is the predicted reserve for LoB $\ell$ and $R^{\ell}$ is the corresponding observed reserve.

\begin{table}[H]
\centering
\caption{Point estimates of unpaid loss reserves for Company 1. Here, $R_1$ and $R_2$ denote reserves for LoB 1 and LoB 2, respectively, and $R=R_1+R_2$ denotes the total reserve. The pGCMM variants differ according to whether LASSO regularization is applied to accident year effects, development year effects, or both.}
\label{tab:reserve_point_estimates}
\begin{tabular}{lrrr}
\hline
Model & LoB 1, $R_1$ & LoB 2, $R_2$ & Total, $R$ \\
\hline
pGCMM (penalize AY and DY) & 7,121,190 & 390,795 & 7,511,985 \\
pGCMM  (penalize AY)        & 7,109,006 & 388,928 & 7,497,935 \\
pGCMM  (penalize DY)        & 7,088,786 & 385,955 & 7,474,742 \\
GCMM                        & 7,087,823 & 385,771 & 7,473,594 \\
Copula-FE                  & 7,516,803 & 427,449 & 7,944,251 \\
Silo-GLMM                    & 7,234,009 & 378,457 & 7,612,466 \\
Actual reserve               & 8,086,094 & 318,380 & 8,404,474 \\
\hline
\end{tabular}
\end{table}

\begin{table}[H]
\centering
\caption{Percentage error of unpaid loss reserve estimates for Company 1. Percentage error is computed as $(\widehat{R}-R)/R \times 100$, where $R$ denotes the observed reserve and $\widehat{R}$ the estimated reserve. Positive values indicate overestimation and negative values indicate underestimation.}
\label{tab:reserve_percentage_error}
\begin{tabular}{lrrr}
\hline
Model & Personal Auto & Commercial Auto & Total \\
\hline
pGCMM (penalize AY and DY)  & -11.93\% & 22.74\% & -10.62\% \\
pGCMM  (penalize AY)        & -12.08\% & 22.16\% & -10.79\% \\
pGCMM  (penalize DY)         & -12.33\% & 21.22\% & -11.06\% \\
GCMM                         & -12.35\% & 21.17\% & -11.08\% \\
Copula-FE                     & -7.04\%  & 34.26\% & -5.48\% \\
Silo-GLMM                     & -10.54\% & 18.87\% & -9.42\% \\
\hline
\end{tabular}
\end{table}

The results show that all models underestimate the total reserve for Company~1, but the allocation of reserve error across lines differs materially. The Copula-FE benchmark provides the closest total reserve point estimate among the models considered. This should not be interpreted as uniform dominance in reserving performance, however, because the aggregate accuracy is accompanied by a relatively large overestimation of the commercial auto reserve. In contrast, the GCMM and pGCMM produce more stable line-specific reserve estimates and substantially reduce the overestimation of the commercial LoB. This distinction is important actuarially: an apparently accurate aggregate reserve may still mask material line-level misallocation, which can affect portfolio monitoring, capital allocation, and line-of-business management.

The comparison also shows that introducing random effects and sparsity changes the allocation of reserves across LoBs. The mixed-model structure borrows information across companies while preserving company-level heterogeneity, whereas the LASSO penalty stabilizes accident-year and development-year effects in the sparse tail of the triangles. Among the penalized specifications, the model penalizing both accident-year and development-year effects provides the most parsimonious specification and is therefore used for the subsequent bootstrap and risk-capital analysis.

\begin{table}[H]
\centering
\caption{Bootstrap summary of total reserve estimates for Company 1. Bias is computed as the percentage difference between the bootstrap mean and the reserve estimate, and CV denotes the coefficient of variation, defined as the ratio of the bootstrap standard deviation to the reserve estimate}
\label{tab:bootstrap_total_reserve}
\begin{tabular}{lrrrrr}
\hline
Model & Reserve & Bootstrap mean & Bias & Std. dev. & CV \\
\hline
pGCMM   & 7,511,985 & 7,515,040 & -0.04\% & 523,765 & 0.070 \\
GCMM    & 7,473,594 & 7,474,344 & -0.01\% & 525,478 & 0.070 \\
Copula-FE & 7,944,251 & 7,922,295 & 0.28\% & 589,482 & 0.074 \\
Silo-GLMM & 7,612,466 & 7,619,048 & -0.09\% & 685,020 & 0.090 \\
\hline
\end{tabular}
\end{table}

\begin{figure}[H]
    \centering
    \begin{subfigure}{0.32\textwidth}
        \centering
        \includegraphics[width=\linewidth]{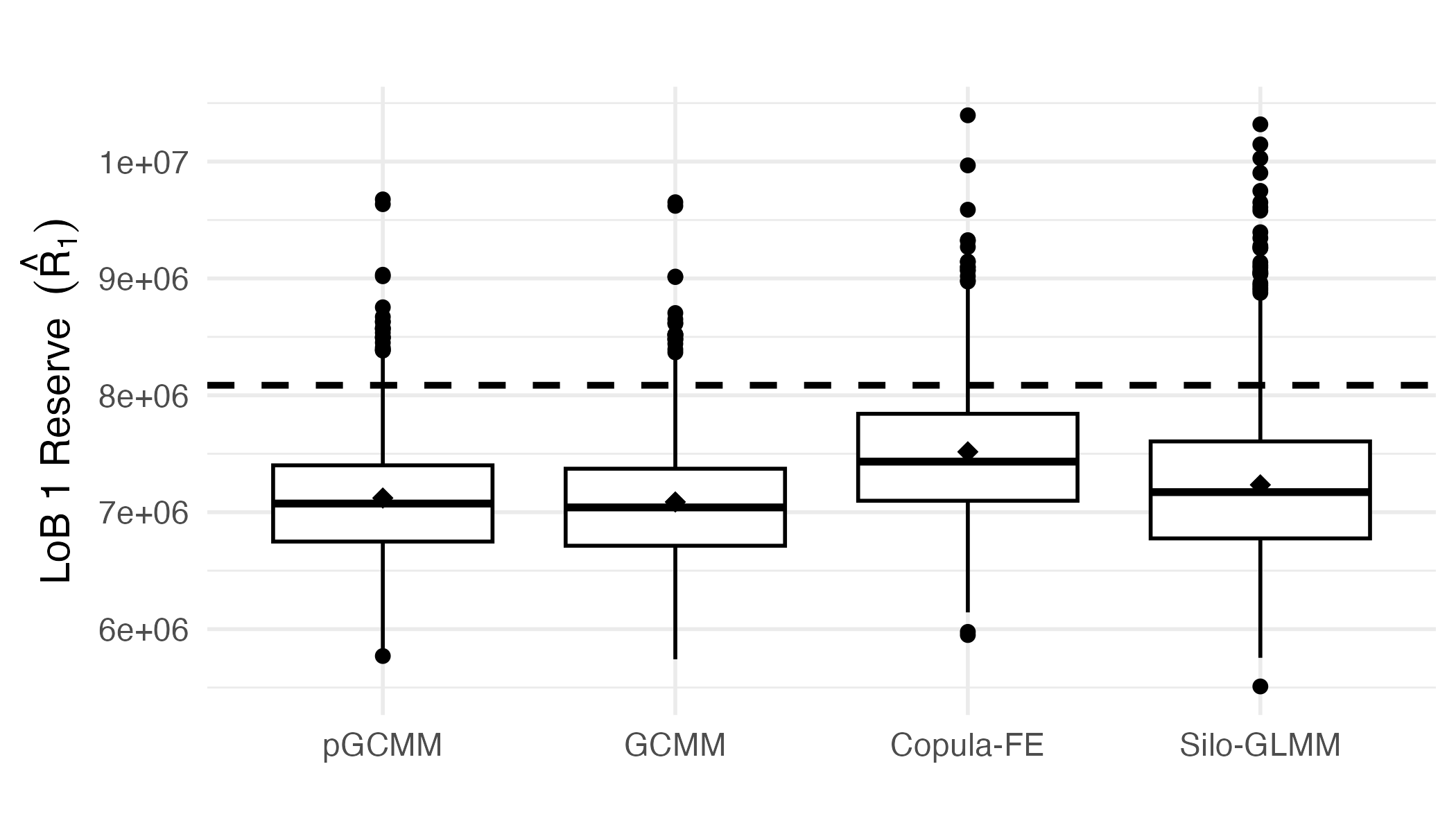}
        \caption{Personal auto}
    \end{subfigure}
    \begin{subfigure}{0.32\textwidth}
        \centering
        \includegraphics[width=\linewidth]{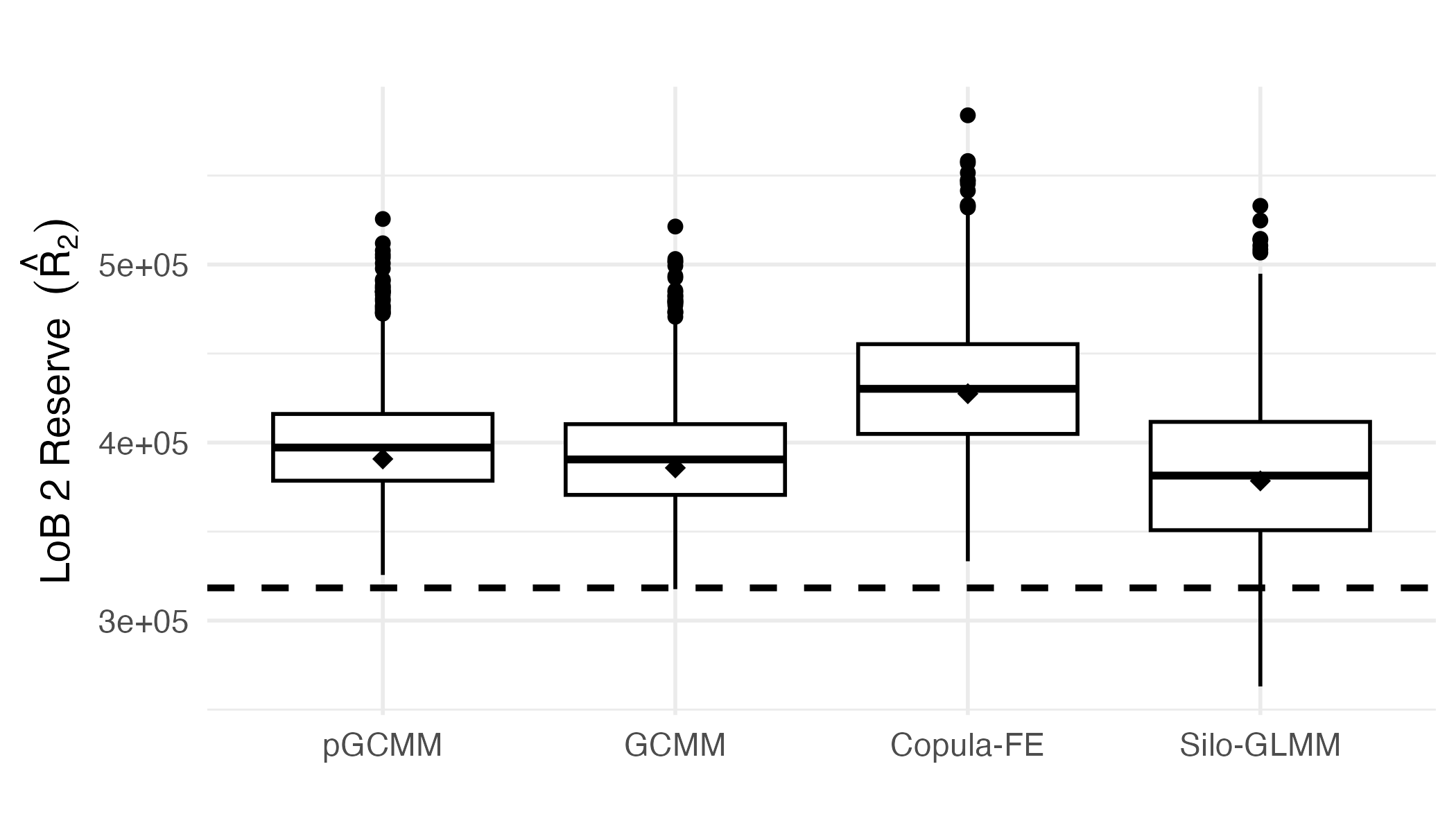}
        \caption{Commercial auto}
    \end{subfigure}
    \begin{subfigure}{0.32\textwidth}
        \centering
        \includegraphics[width=\linewidth]{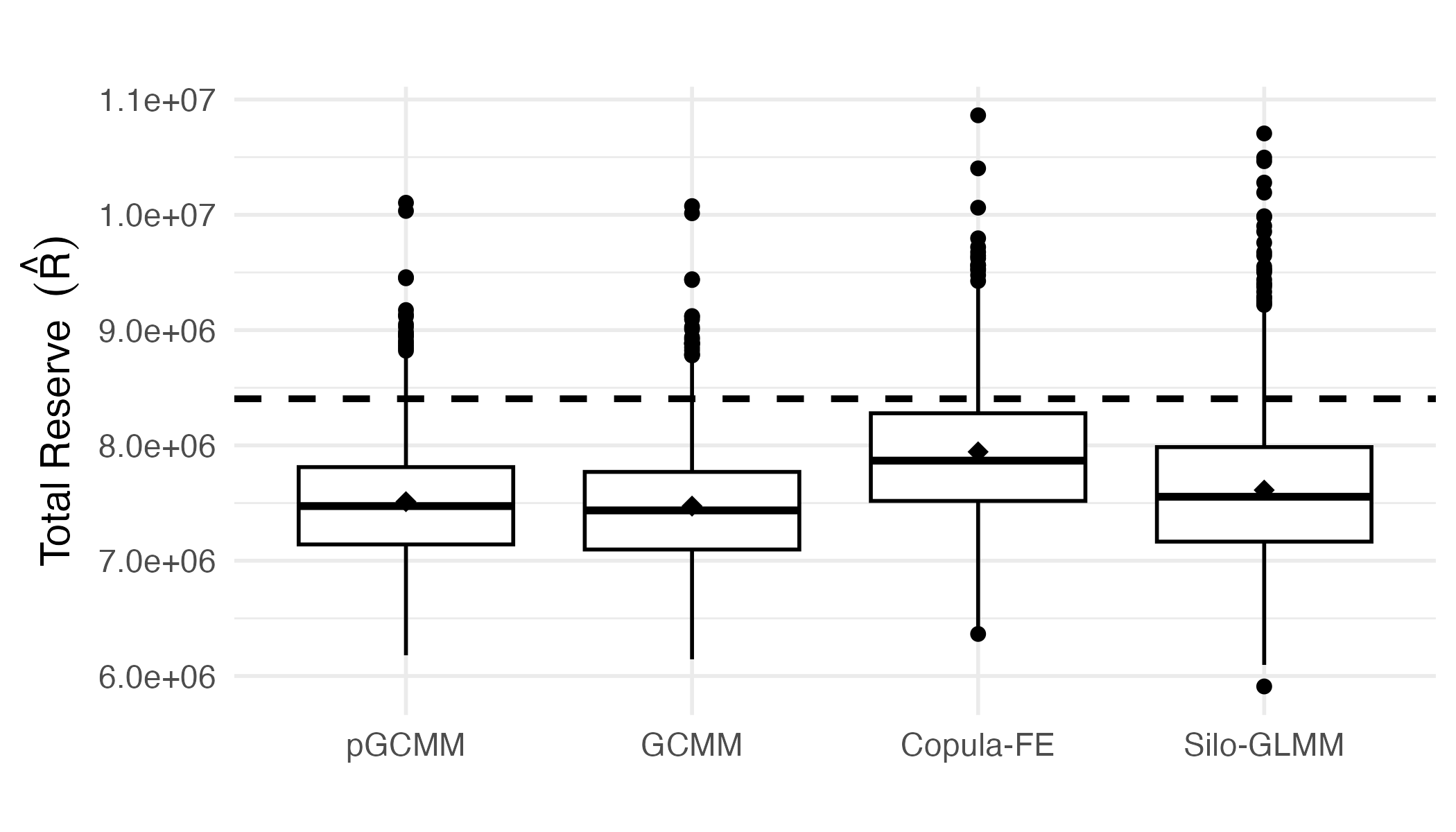}
        \caption{Total reserve}
    \end{subfigure}
    \caption{Bootstrap distributions of unpaid loss reserve estimates for Company 1. Boxes summarize the bootstrap distributions, diamonds denote the corresponding reserve point estimates, and the dashed vertical line denotes the observed reserve.\color{black}}
    \label{fig:bootstrap_distribution_comparison}
\end{figure}

To assess the uncertainty around the point estimates, we apply the modified bootstrap procedure described in Algorithm~\ref{alg:pgcmm-bootstrap}. The bootstrap analysis provides the predictive distribution of the total reserve under each model and allows us to compare the stability of the resulting reserve estimates.

Table~\ref{tab:bootstrap_total_reserve} summarizes the bootstrap distribution of the total reserve. All four models exhibit small bootstrap bias, indicating that the point estimates are broadly aligned with their corresponding bootstrap means. However, the variability differs across models. The Silo-GLMM produces the largest bootstrap standard deviation and coefficient of variation, reflecting the loss of diversification information when the two LoBs are modeled separately. The Copula-FE reduces this variability, while the mixed-effect copula models produce the lowest coefficients of variation. The pGCMM retains essentially the same coefficient of variation as the GCMM, while using a more parsimonious fixed-effect structure. This supports the role of penalization as a stabilizing device rather than merely a variable-selection mechanism.

We examine further the uncertainty associated with reserve prediction.  Figure~\ref{fig:bootstrap_distribution_comparison} displays the bootstrap distributions of reserves for the personal auto, commercial auto, and total portfolio reserves. The horizontal line represents the actual reserve, and the diamond shape within the box represents the corresponding point estimates. The figure illustrates that the Silo-GLMM model has the widest predictive distributions, especially for the total reserve, while the GCMM and pGCMM-based reserve distributions are more concentrated. This is consistent with the numerical bootstrap summaries and reinforces the value of modeling residual cross-LoB dependence after accounting for insurer-level heterogeneity and penalization.


Table~\ref{tab:tvar_estimates} reports TVaR values for the predictive distribution of the total reserve. 
For the Silo-GLMM, the portfolio TVaR is computed as the sum of the line-specific TVaRs, corresponding to a conservative no-diversification aggregation. 
For Copula-FE, GCMM, and pGCMM, TVaR is computed directly from the joint predictive distribution of the total reserve. The comparison further allows us to assess the effects of modeling insurer heterogeneity through random effects and of penalization through LASSO regularization on tail risk estimates. 

Table~\ref{tab:risk_capital} reports the incremental tail capital, defined here as the excess of TVaR at level $\alpha$ over TVaR$_{60}$,
\begin{equation}
    RC_{\alpha}
    =
    \text{TVaR}_{\alpha}
    -
    \text{TVaR}_{60},
    \qquad 
    \alpha \in \{80\%,85\%,90\%,95\%,99\%\}.
\end{equation}
This normalization focuses the comparison on the additional capital required when moving from a moderate-tail to a more extreme-tail risk measure. 
Table~\ref{tab:risk_capital_gain} then reports the percentage reduction in this incremental tail capital relative to the Silo-GLMM.

\begin{table}[H]
\centering
\caption{TVaR estimates for the total reserve for Company 1.}
\label{tab:tvar_estimates}
\begin{tabular}{lrrrrrr}
\hline
Model & TVaR$_{60}$ & TVaR$_{80}$ & TVaR$_{85}$ & TVaR$_{90}$ & TVaR$_{95}$ & TVaR$_{99}$ \\
\hline
pGCMM   & 8,016,040 & 8,296,933 & 8,403,554 & 8,553,519 & 8,794,819 & 9,353,323 \\
GCMM   & 7,976,621 & 8,257,105 & 8,365,286 & 8,515,486 & 8,761,293 & 9,326,912 \\
Copula-FE & 8,494,293 & 8,799,525 & 8,914,652 & 9,074,084 & 9,319,283 & 9,891,141 \\
Silo-GLMM & 8,312,683 & 8,704,288 & 8,863,507 & 9,082,936 & 9,452,114 & 10,281,955 \\
\hline
\end{tabular}
\end{table}

\begin{table}[H]
\centering
\caption{Risk capital estimates relative to TVaR$_{60}$ for Company 1.}
\label{tab:risk_capital}
\begin{tabular}{lrrrrr}
\hline
Model & RC$_{80}$ & RC$_{85}$ & RC$_{90}$ & RC$_{95}$ & RC$_{99}$ \\
\hline
pGCMM   & 280,894 & 387,514 & 537,479 & 778,779 & 1,337,284 \\
GCMM     & 280,485 & 388,666 & 538,865 & 784,672 & 1,350,291 \\
Copula-FE & 305,232 & 420,359 & 579,790 & 824,989 & 1,396,848 \\
Silo-GLMM & 391,606 & 550,824 & 770,253 & 1,139,431 & 1,969,272 \\
\hline
\end{tabular}
\end{table}\begin{table}[H]
\centering
\caption{Risk capital gains relative to the Silo-GLMM benchmark.}
\label{tab:risk_capital_gain}
\begin{tabular}{lrrrrr}
\hline
Model & RC$_{80}$ & RC$_{85}$ & RC$_{90}$ & RC$_{95}$ & RC$_{99}$ \\
\hline
pGCMM  & 28.3\% & 29.6\% & 30.2\% & 31.7\% & 32.1\% \\
GCMM     & 28.4\% & 29.4\% & 30.0\% & 31.1\% & 31.4\% \\
Copula-FE & 22.1\% & 23.7\% & 24.7\% & 27.6\% & 29.1\% \\
Silo-GLMM & 0.0\% & 0.0\% & 0.0\% & 0.0\% & 0.0\% \\
\hline
\end{tabular}
\end{table}

The results show that explicitly modeling dependence across LoBs leads to materially lower risk capital than the silo benchmark. 
The Silo-GLMM produces the largest incremental capital at all tail levels because the two LoBs are modeled separately and diversification is not explicitly captured. 
The copula-based models reduce required capital, with the GCMM and pGCMM producing the largest gains relative to the Silo-GLMM. 
At the 99\% TVaR level, the pGCMM reduces incremental capital by approximately 32.1\% relative to the Silo-GLMM, compared with 29.1\% for the Copula-FE. 
These results indicate that accounting jointly for company heterogeneity, residual cross-LoB dependence, and penalized fixed effects improves the stability of the predictive distribution and the resulting capital assessment.

Taken together, the point-estimate and bootstrap results highlight a trade-off between company-specific point accuracy and stability of the full predictive reserve distribution. While Copula-FE is competitive for the total point reserve of Company~1, the mixed-effect copula models yield lower bootstrap variability and lower incremental tail capital. This supports the interpretation of GCMM and pGCMM as credibility-type extensions of copula regression: they borrow information across companies, reduce sensitivity to company-specific fixed effects, and provide a more stable basis for capital assessment.

\begin{figure}[H]
    \centering
    \includegraphics[width=0.5\textwidth]{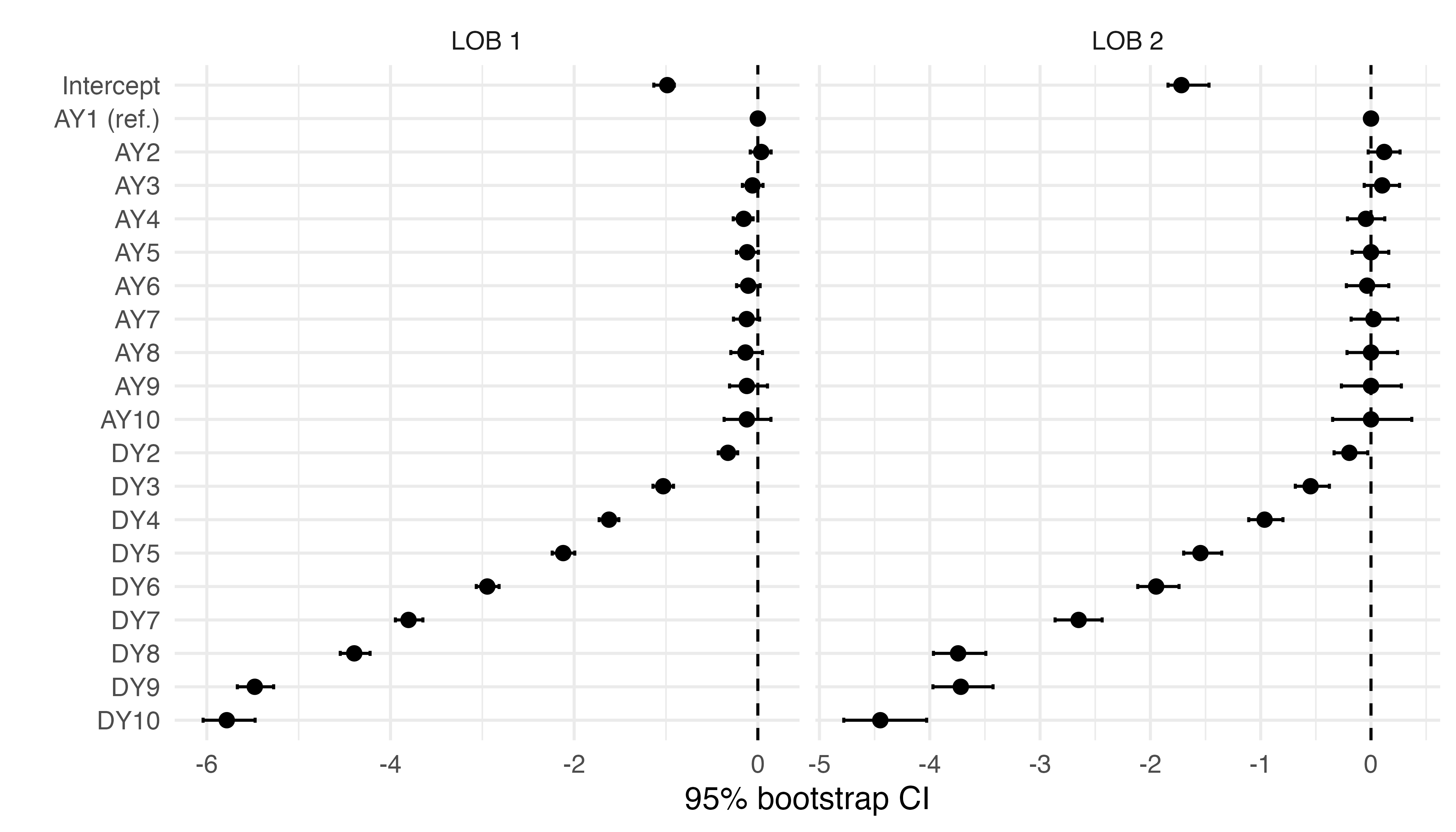}
    \caption{Point estimates and 95\% bootstrap confidence intervals for the fixed-effect parameters in LoB 1 and LoB 2 obtained from the fitted pGCMM. The dashed vertical line indicates zero.}
    \label{fig:beta_bootstrap_CI_sSURCMM_Allasso}
\end{figure}
\begin{figure}[H]
    \centering
    \includegraphics[width=0.5\textwidth]{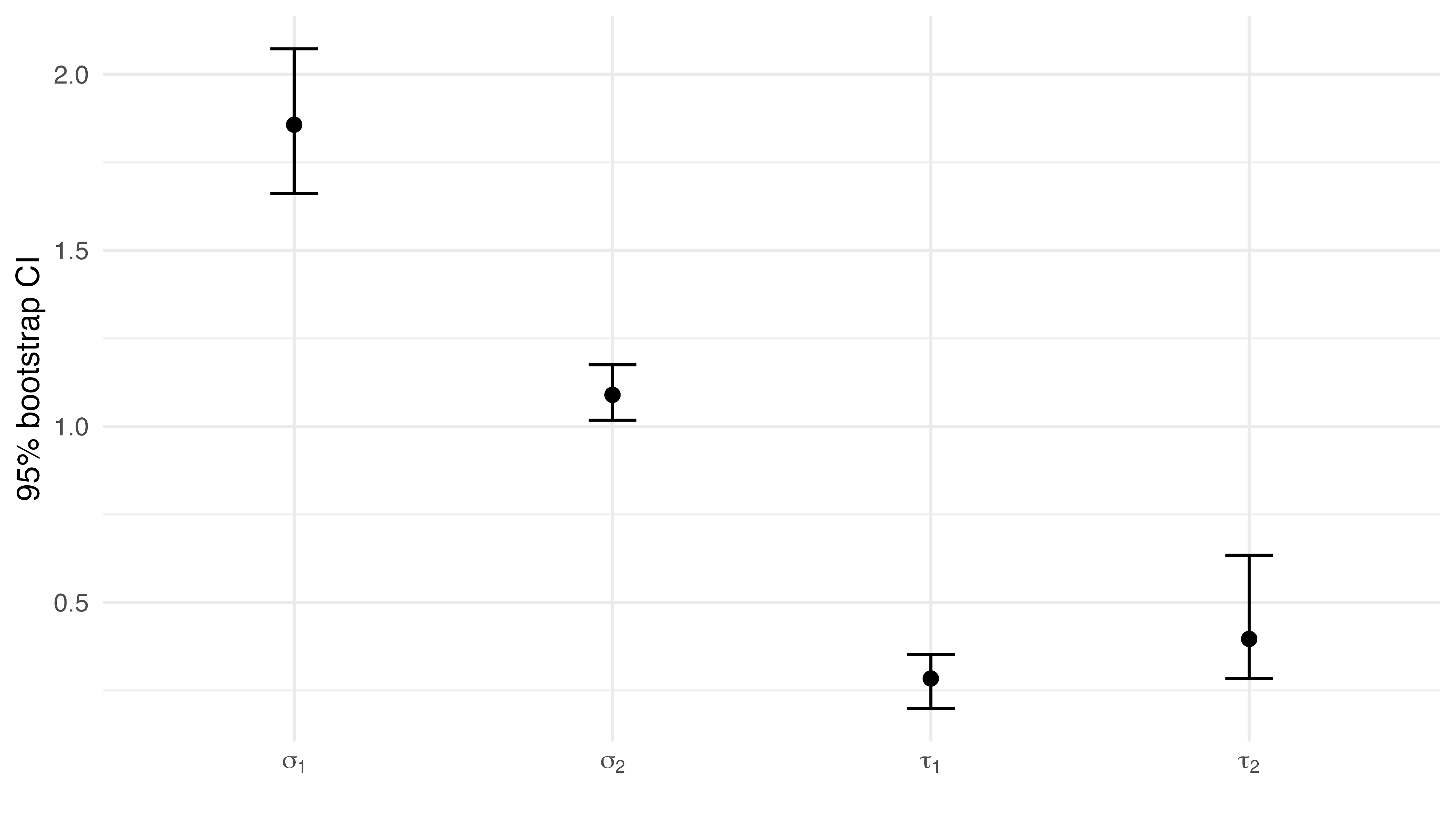}
    \caption{Point estimates and 95\% bootstrap confidence intervals for the gamma shape and random effect standard deviation for LoB 1 and LoB 2 obtained from the fitted pGCMM. The dashed vertical line indicates zero.}
    \label{fig:variance_shape_bootstrap_CI_sSURCMM_Allasso}
\end{figure}
\begin{figure}[H]
    \centering
    \includegraphics[width=0.5\textwidth]{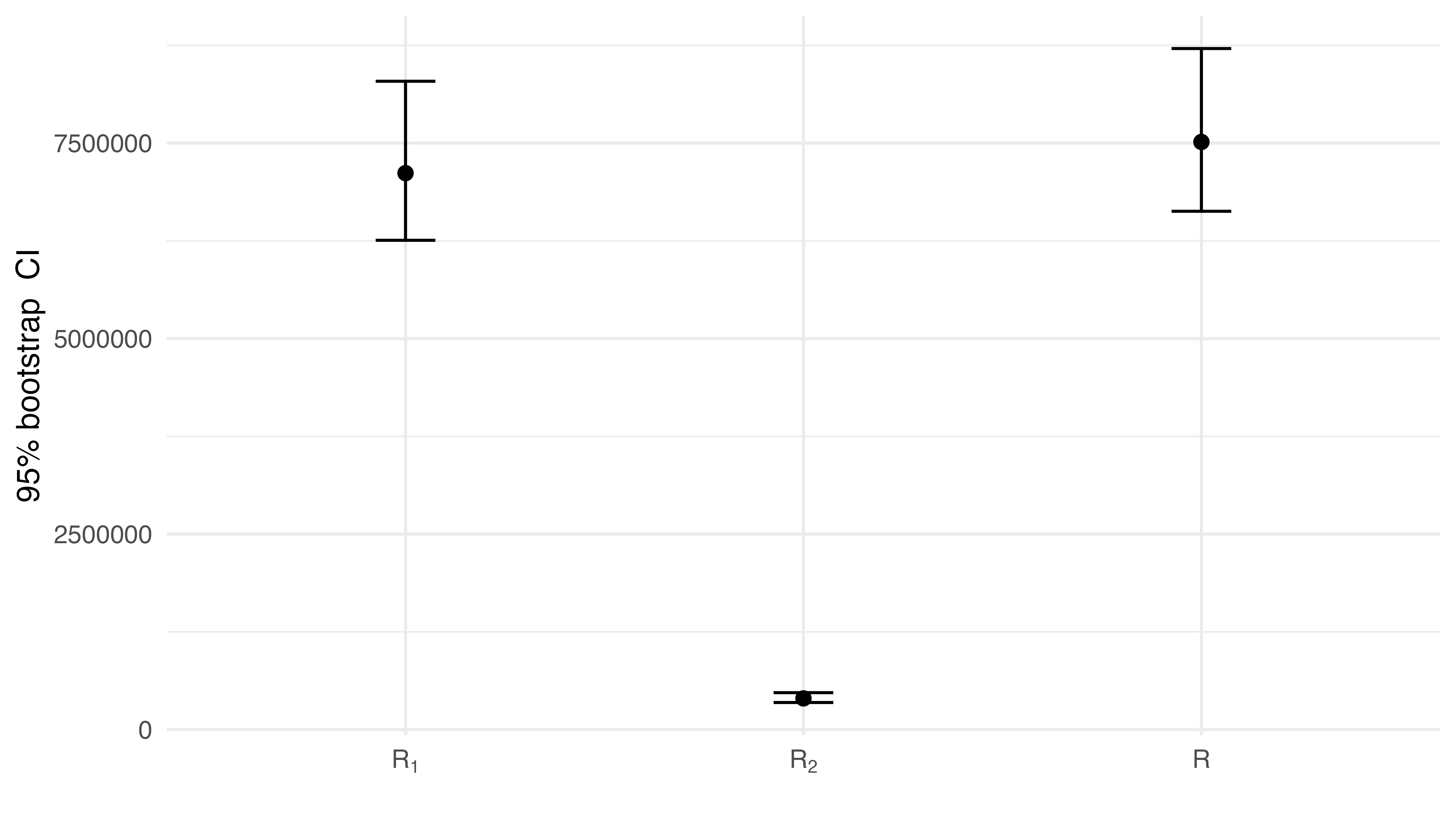}
    \caption{Point estimates and 95\% bootstrap confidence intervals for the reserves of LoB~1 ($R_1$), LoB~2 ($R_2$), and the total reserve ($R$) for Company~1 under the fitted pGCMM.}
    \label{fig:reserve_bootstrap_CI_sSURCMM_Allasso}
\end{figure}
\begin{figure}[H]
    \centering
    \includegraphics[width=0.5\textwidth]{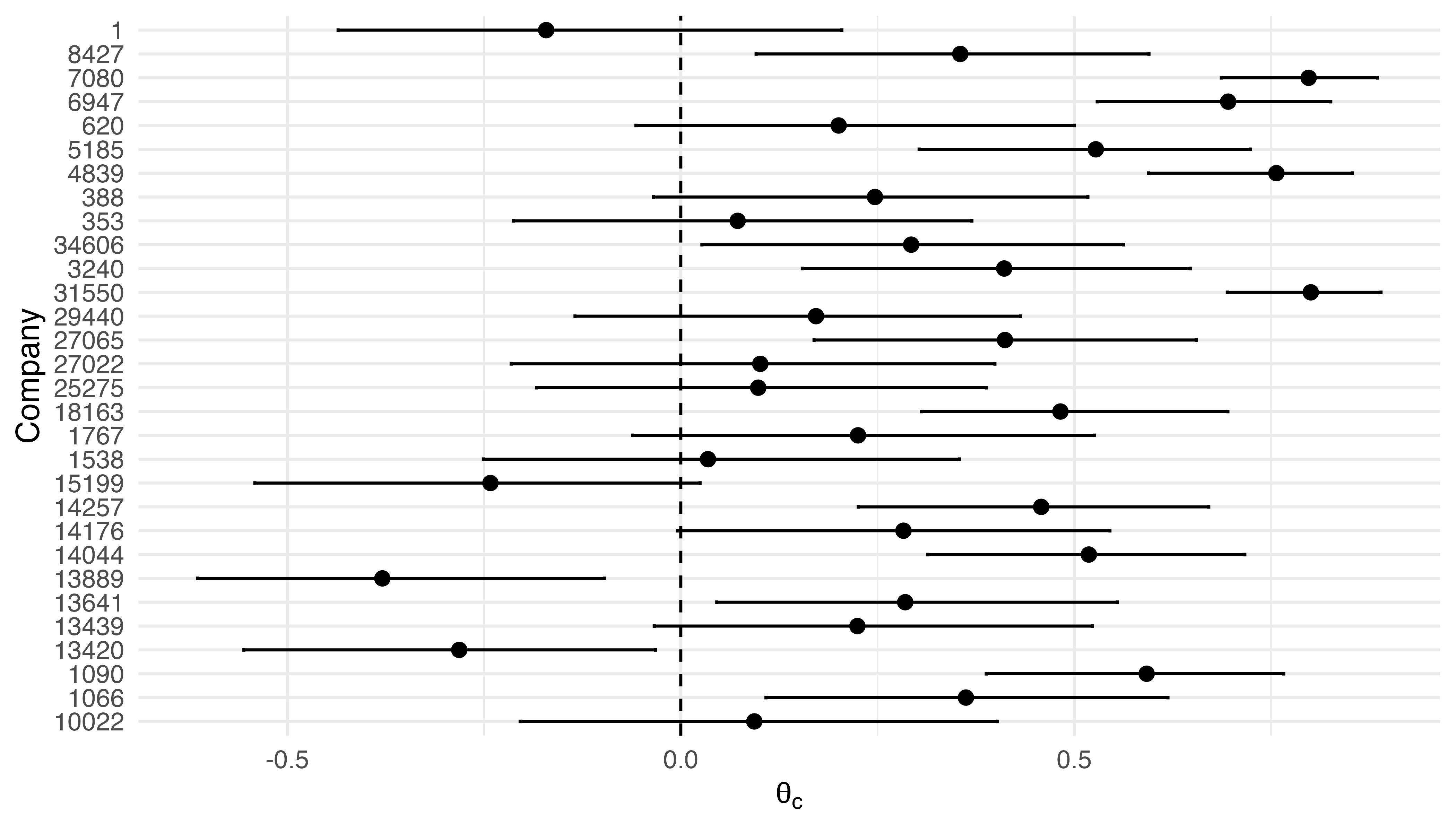}
    \caption{Point estimates and 95\% bootstrap confidence intervals for the company-specific copula dependence parameters $\theta_c$ obtained from the fitted pGCMM. The dashed vertical line indicates zero dependence.}
    \label{fig:theta_bootstrap_CI_sSURCMM_Allasso}
\end{figure}

We next investigate the parameter estimates underlying the fitted pGCMM. Figure~\ref{fig:beta_bootstrap_CI_sSURCMM_Allasso} displays the bootstrap confidence intervals for the fixed-effect parameters, Figure~\ref{fig:variance_shape_bootstrap_CI_sSURCMM_Allasso} summarizes the gamma shape and random-effect standard deviation parameters, Figure~\ref{fig:reserve_bootstrap_CI_sSURCMM_Allasso} reports the resulting reserve estimates, and Figure~\ref{fig:theta_bootstrap_CI_sSURCMM_Allasso} presents the company-specific copula dependence estimates. Corresponding results for the GCMM are provided in Figures~\ref{fig:beta_bootstrap_CI_GCMM_nolasso}, \ref{fig:variance_shape_bootstrap_CI_GCMM_nolasso}, \ref{fig:theta_bootstrap_CI_GCMM_nolasso}, and \ref{fig:reserve_bootstrap_CI_GCMM_Allasso}. The two models yield broadly similar reserve and dependence estimates, with the main differences appearing in the confidence intervals of the fixed effects. This suggests that the primary role of penalization is to stabilize fixed-effect estimation and improve parsimony, leading to more consistent reserve uncertainty and risk-capital assessments.

Most companies exhibit positive dependence between the two lines of business, although the strength of dependence varies substantially across companies. For Company 1, the estimated dependence parameter is negative, consistent with previous analyses \citep{Shi2011}. However, the corresponding bootstrap confidence interval contains zero, indicating substantial estimation uncertainty and suggesting that the evidence for negative dependence is weak. This uncertainty may arise from the limited amount of company-specific information available for estimating dependence, particularly when the marginal variability in one or both lines of business is large compared to the LoB pairwise dependence.

For Company 1, another possible reason for the weak or negative estimate is that the current specification does not explicitly model serial dependence across development years within each line. Any remaining temporal structure in the residuals may affect the pseudo-observations used for copula estimation. We therefore interpret Company~1's estimate as evidence of weak residual dependence rather than strong negative dependence.

\section{Simulation Study}

The real-data analysis illustrates the practical implications of the proposed framework for reserve prediction and risk capital assessment. However, because the true parameter values and reserves are unknown, it precludes a direct assessment of estimation accuracy. We therefore conduct a simulation study to evaluate the finite-sample performance of the proposed estimation procedure under three settings. 

In Setting 1, the simulated data are generated from the fitted $L_{1}$-penalized generalized copula mixed model (pGCMM). This setting evaluates whether the proposed estimator can correctly recover the nonzero coefficients while shrinking the truly zero coefficients toward zero. We begin with the fitted pGCMM obtained from the real data analysis in Section \ref{application}. Then, the estimated fixed effects, variance of the random effects, gamma shape, and company-specific copula parameters are treated as the actual model parameters. The accident years (AY) 5, 8, 9, and 10 effects in LoB 2 are set to zero. All development year (DY) effects are set to non-zero.

The company-specific random effects are generated independently as $b_c^{(1)} \sim N(0,\tau_1^2)$ and $b_c^{(2)} \sim N(0,\tau_2^2)$, where $\tau_1$ and $\tau_2$ are the variance estimates from the fitted model. The systematic components are then computed as $\eta_{ijc}^{(\ell)} =\mathbf{x}_{ijc}^{(\ell)\top}\boldsymbol{\beta}^{(\ell)}+b_c^{(\ell)}, \ell=1,2,$ with corresponding mean loss ratios $\mu_{ijc}^{(\ell)} = \exp\left(\eta_{ijc}^{(\ell)}\right)$.

Dependence between the two LoBs is introduced through company-specific Gaussian copulas. For company $c$, copula coordinates $\left(U_{ijc}^{(1)},U_{ijc}^{(2)}\right) \sim C(\cdot,\cdot;\theta_c)$ are generated from a Gaussian copula with parameter $\theta_c$, where $\theta_c$ is the estimated company-specific dependence parameter obtained from the fitted pGCMM. The copula coordinates are then transformed to loss ratios through the gamma inverse cumulative distribution function (CDF). 
$$Y_{ijc}^{(\ell)}
=
F_{\ell}^{-1}
\left(
u_{ijc}^{(\ell)};
\sigma_\ell,
\frac{\mu_{ijc}^{(\ell)}}{\sigma_\ell}
\right),
\qquad \ell=1,2,$$ where $\sigma_\ell$ is the estimated gamma shape parameter, and the corresponding scale parameter is
$\dfrac{\mu_{ijc}^{(\ell)}}{\sigma_\ell}$. Thus, $E\left[Y_{ijc}^{(\ell)}\mid b_c^{(\ell)}\right] =
\mu_{ijc}^{(\ell)}$ and $\mathrm{Var}\left[Y_{ijc}^{(\ell)}\mid b_c^{(\ell)}\right] = \dfrac{\mu_{ijc}^{(\ell)2}}{\sigma_\ell}.$

In Setting 2, the simulated data are generated from the fitted unpenalized generalized copula mixed model (GCMM). This evaluates the impact of $L_{1}$ penalization when the true model fixed effects are not exactly zero without adversely affecting reserve estimation and risk capital calculations. We begin with the fitted GCMM obtained from the real data analysis, where no $L_{1}$-penalty is imposed. Unlike Setting 1, the accident years 5, 8, 9, and 10 effects in LoB 2 are small but nonzero. 

In Setting 3, the simulated data are generated from the pGCMM with heavy-tailed random effects. This evaluates the robustness of the proposed estimator to departures from the Gaussian random effect assumption. As in Setting 1, the estimated fixed effects, gamma shape parameters, and company-specific copula parameters obtained from the fitted pGCMM are treated as the actual model parameters. The accident years 5, 8, 9, and 10 effects in LoB 2 are set to zero.

The data-generating mechanism is identical to that of Setting 1 except for the distribution of the company-specific random effects. We generate $b_c^{(\ell)}
= \tau_2 \sqrt{\frac{\nu-2}{\nu}},T_c^{(2)},$ where $T_c^{(1)},T_c^{(2)} \stackrel{\text{i.i.d.}}{\sim} t_{\nu}$ are Student's-t distribution with degrees of freedom $\nu=3$. The scaling factor $\sqrt{\dfrac{(\nu-2)}{\nu}}$ is introduced so that $\tau_\ell^2, \ell=1,2,$ equals the variance of the Gaussian random effects used in Setting 1 while allowing for substantially heavy-tails.

For each setting, we independently generate $R=100$ datasets, referred to as simulation runs. Each simulation run consists of 30 pairs of loss triangles corresponding to the two LoBs and preserving the accident year, development year, company, and premium observed in the real dataset.

For every simulated dataset, the proposed pGCMM is fitted using the iterative estimation procedure described in Algorithm~\ref{alg:pgcmm-estimation}. To select the $L_{1}$-penalty parameters, a grid search is performed over candidate penalty values for the two LoBs. The optimal penalty combination is chosen using the Akaike Information Criterion (AIC). Following model selection, the final parameter estimates, company-specific copula parameters, reserve estimates, and risk capital measures are obtained from the selected model.

The simulation results are evaluated by examining the finite-sample properties of the estimators and reserve estimates across 100 independent simulation runs. For each parameter, the Monte Carlo sampling distribution of the estimates is compared with the corresponding actual value to assess variability and concentration around the true parameter. Reserve performance is assessed using the relative reserve error, $RE=\frac{\widehat{R}-R}{R}$, where $R$ and $\widehat{R}$ denote the actual and estimated reserves, respectively. Since the lower triangles are generated as part of the simulation setting, the actual reserves are known and are computed by aggregating the simulated lower-triangle losses across accident years and companies. The distribution of the relative reserve errors across simulation runs is then used to assess the accuracy and stability of reserve estimation under each simulation setting.

\begin{figure}[ht]
    \centering
    \includegraphics[width=0.8\textwidth]{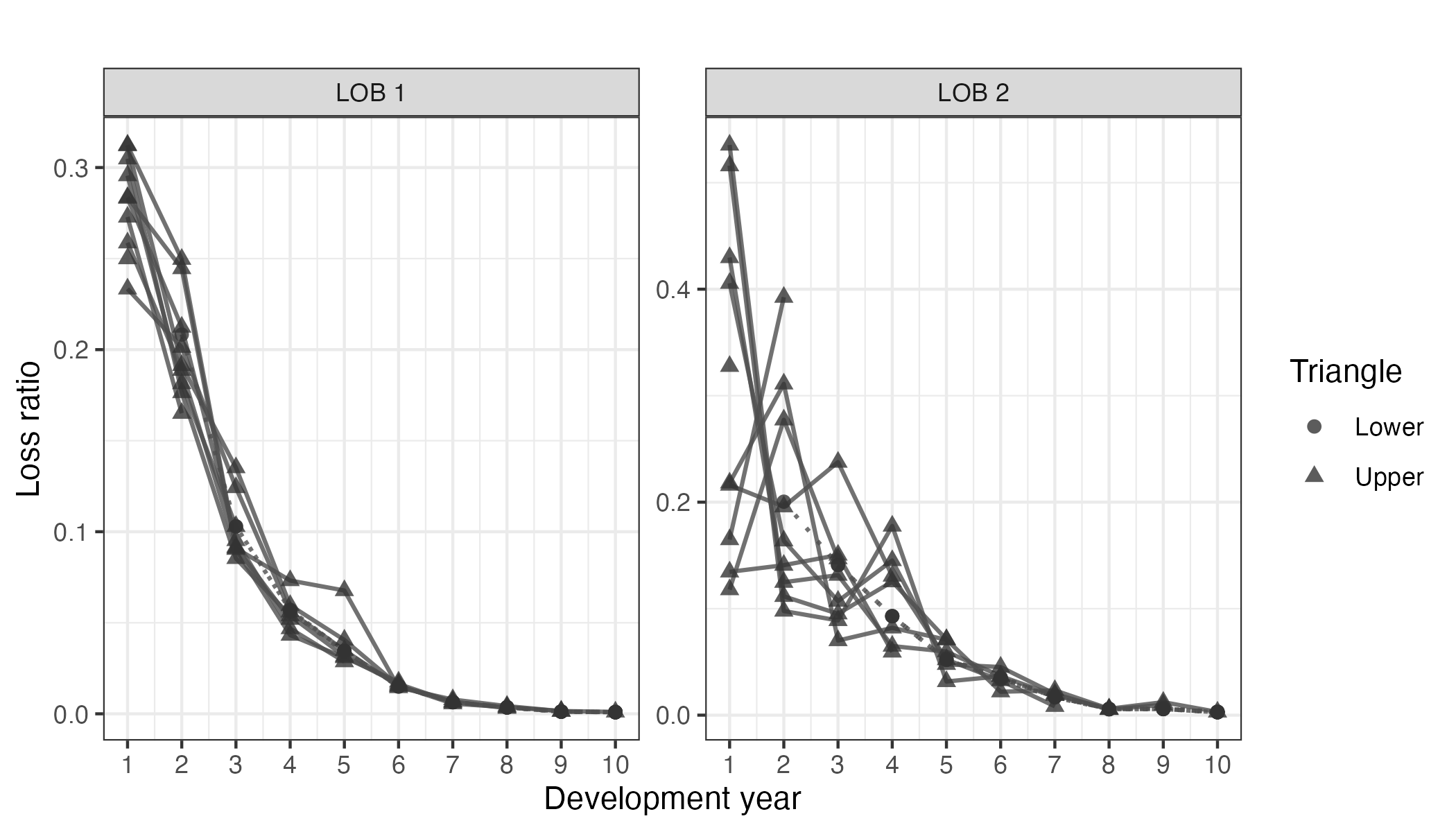}
    \caption{One simulated realization from Setting 1 for Company 1. Each trajectory corresponds to an accident year. Triangles denote observations from the simulated upper triangle, while circles denote observations from the simulated lower triangle.}
    \label{fig:simulated_triangle_company1}
\end{figure}

Figure \ref{fig:simulated_triangle_company1} displays one simulated realization from Setting 1 for Company 1. The loss ratios in LoB 1 exhibit a relatively smooth decline over development years, whereas LoB 2 displays substantially greater variability and more irregular development patterns, particularly during the early development periods. This increased volatility suggests that reserve prediction is inherently more challenging for LoB 2 with linear models.

Next, we answer the following questions on evaluating the pGCMM and the proposed estimation procedure: (i) Do the non-zero coefficients recover? (ii) Does LASSO identify actual zero effects? (iii) Does reserve estimation remain robust under random effect distribution misspecification?

\begin{figure}[ht]
    \centering
    \includegraphics[width=0.8\textwidth]{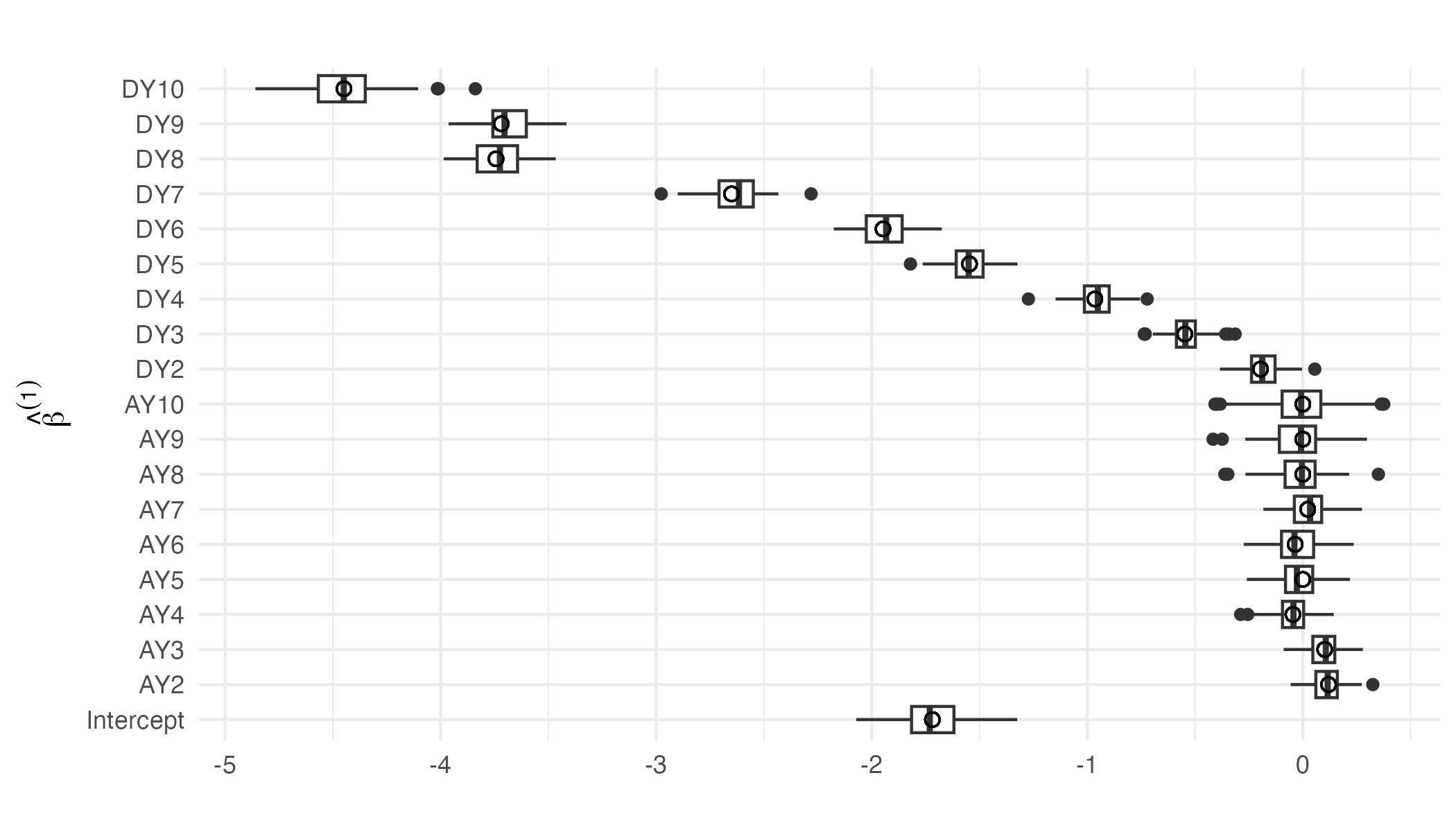}
    \caption{Monte Carlo sampling distributions of the estimated fixed effects for LoB 2 under Setting 1, based on 100 simulation runs. Boxplots summarize the parameter estimates, while open circles denote the actual values.}
    \label{fig:fixed_effects_LOB2_MC_pGCMM}
\end{figure}

Figure \ref{fig:fixed_effects_LOB2_MC_pGCMM} shows the Monte Carlo sampling distributions of the LoB 2 fixed-effect estimates under Setting 1. The development-year effects are accurately recovered with relatively small variability. The accident-year effects AY5, AY8, AY9, and AY10, whose true values are zero, are concentrated at zero. AY7, which corresponds to the smallest nonzero coefficient, occasionally shrinks toward zero, suggesting that the proposed estimator appropriately treats effects of negligible magnitude as candidates for exclusion while retaining the larger accident-year and development-year effects. 

Figures \ref{fig:fixed_effects_LOB2_MC_GCMM} and \ref{fig:fixed_effects_LOB2_MC_pGCMM_b_t3} show the corresponding results under Settings 2 and 3, respectively. Similar estimation behavior is observed across all three settings. Figures \ref{fig:fixed_effects_LOB1_MC_pGCMM}, \ref{fig:fixed_effects_LOB1_MC_GCMM}, and \ref{fig:fixed_effects_LOB1_MC_pGCMM_b_t3} show that the nonzero fixed effects in LoB 1 are recovered under all three simulation settings.

Figures \ref{fig:var_shape_MC_pGCMM}, \ref{fig:var_shape_MC_GCMM}, and \ref{fig:var_shape_MC_pGCMM_b_t3} show Monte Carlo sampling distributions of the estimated gamma shape parameters and random-effect standard deviations under Settings 1, 2, and 3, respectively, based on 100 simulation runs. All four estimator sampling distributions are concentrated around their true values in all settings. The gamma shape parameter estimators exhibit similar variability across the three settings. Greater variability is observed for the random effect standard deviation estimators, particularly $\hat{\tau}_2$. The largest variability occurs under Setting 3, where several extreme estimates are observed, reflecting the additional uncertainty introduced by the heavy-tailed random effects. However, the sampling distributions remain centered around the actual values, suggesting that the proposed estimation procedure is reasonably robust to departures from the Gaussian random effect assumption.

\begin{figure}[ht]
    \centering
    \includegraphics[width=0.8\textwidth]{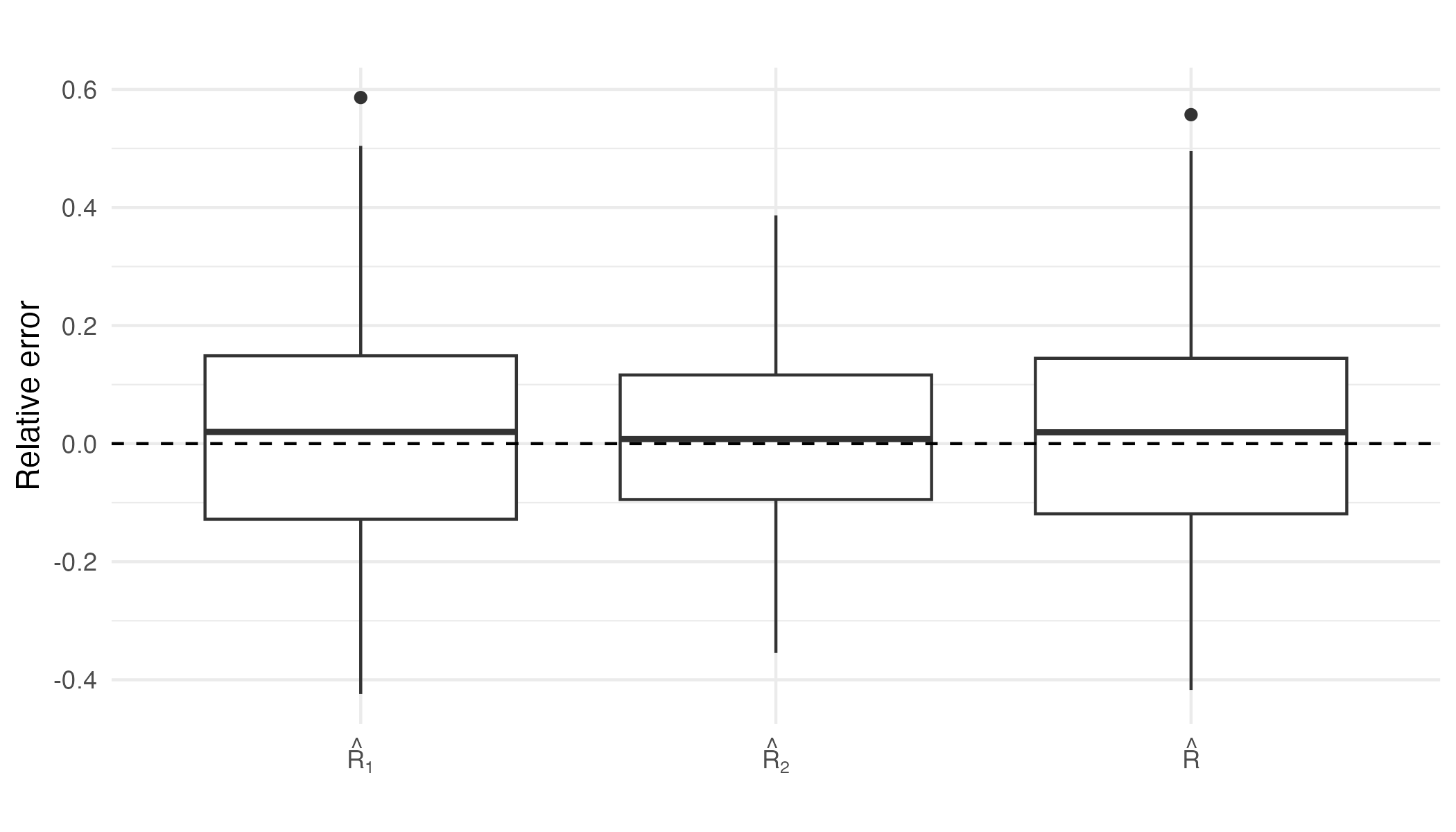}
    \caption{Monte Carlo distributions of the relative reserve errors for LoB 1, LoB 2, and the aggregate reserve under Setting 1, based on 100 simulation runs. The dashed horizontal line corresponds to zero relative error.}
    \label{fig:reserve_rel_error_MC_pGCMM}
\end{figure}

\begin{figure}[ht]
    \centering
    \includegraphics[width=0.8\textwidth]{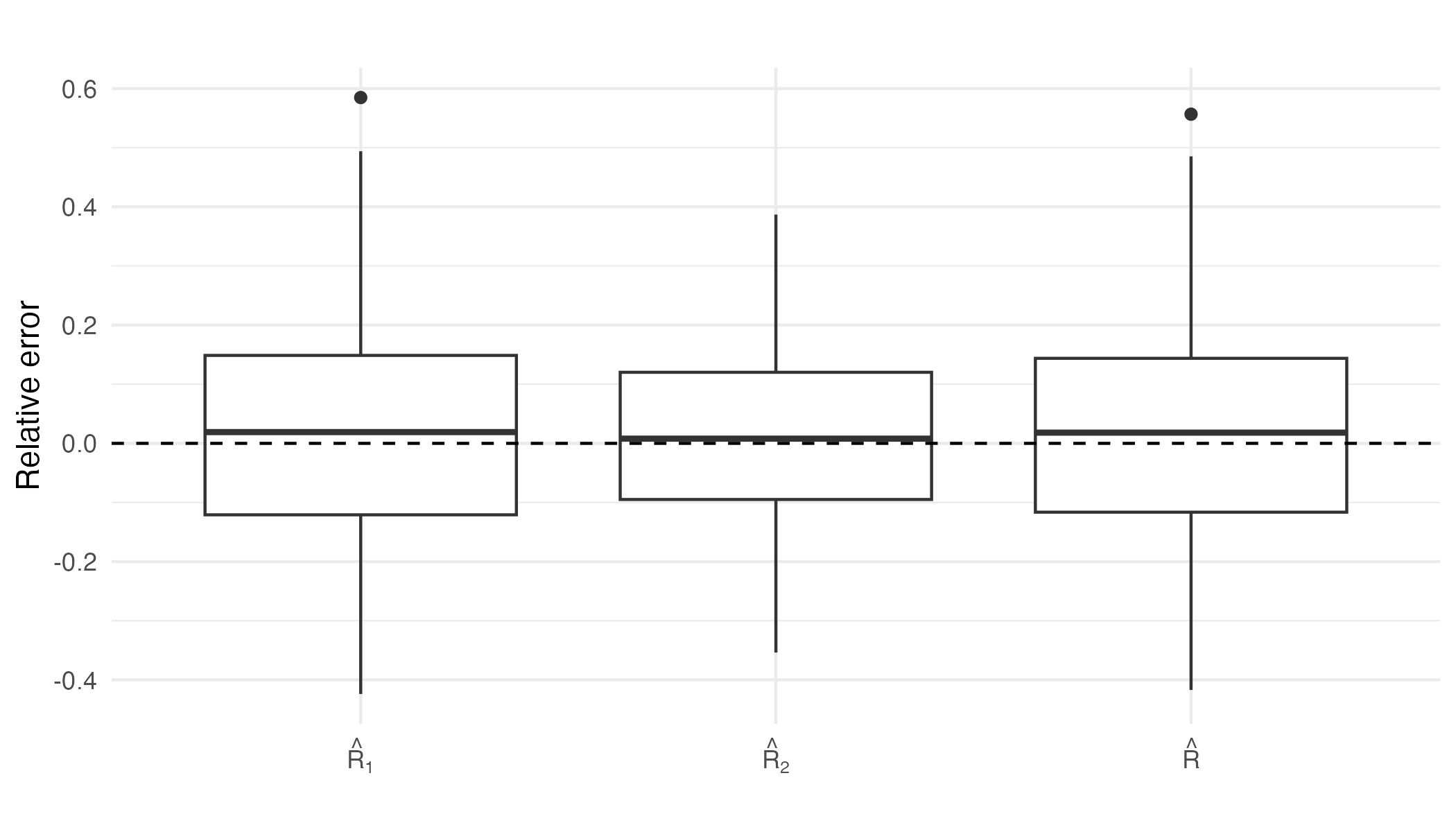}
    \caption{Monte Carlo distributions of the relative reserve errors for LoB 1, LoB 2, and the aggregate reserve under Setting 2, based on 100 simulation runs. The dashed horizontal line corresponds to zero relative error.}
    \label{fig:reserve_rel_error_MC_GCMM}
\end{figure}

\begin{figure}[ht]
    \centering
    \includegraphics[width=0.8\textwidth]{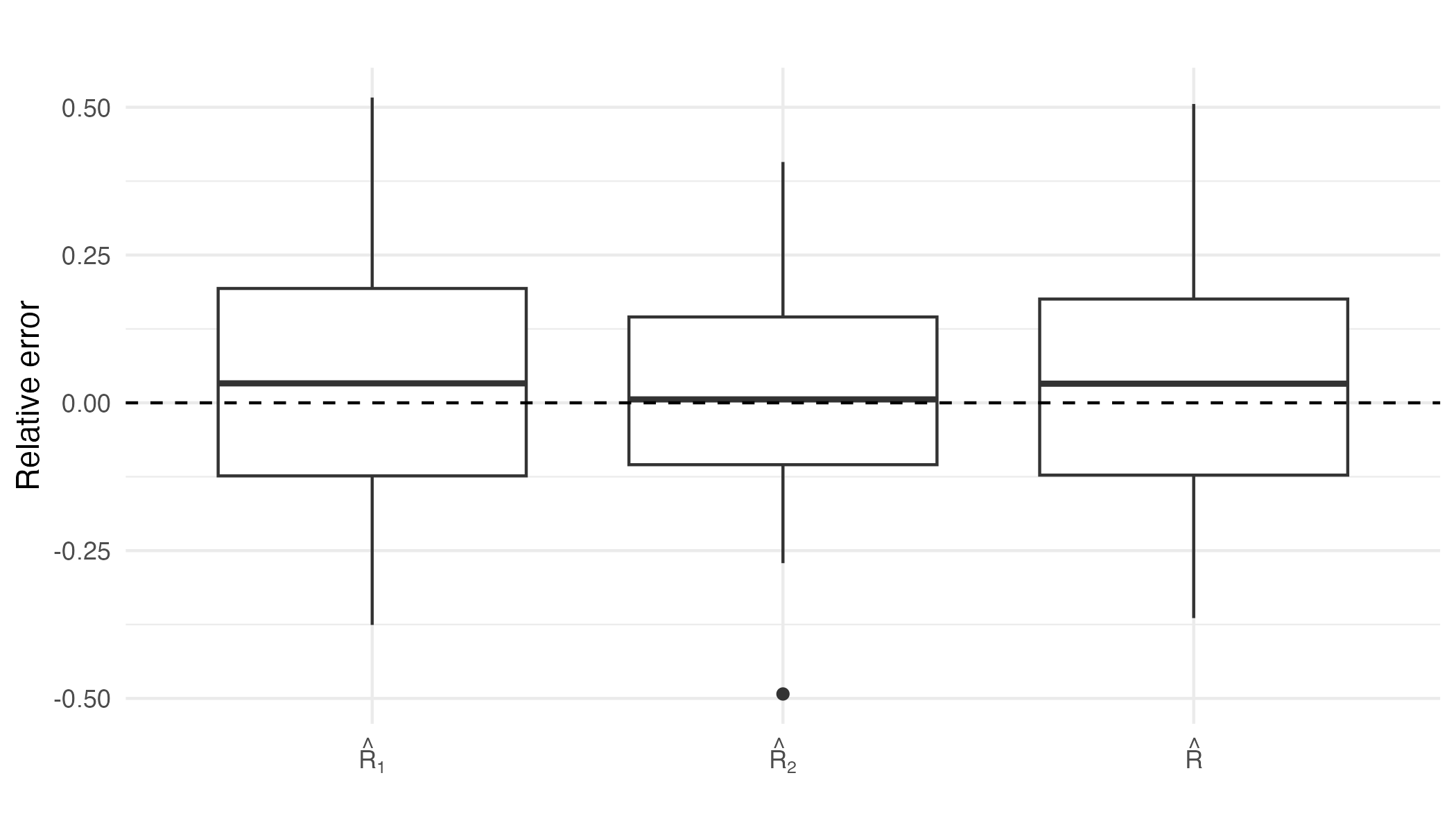}
    \caption{Monte Carlo distributions of the relative reserve errors for LoB 1, LoB 2, and the aggregate reserve under Setting 3, based on 100 simulation runs. The dashed horizontal line corresponds to zero relative error.}
    \label{fig:reserve_rel_error_MC_pGCMM_b_t3}
\end{figure}

Figures \ref{fig:reserve_rel_error_MC_pGCMM}, \ref{fig:reserve_rel_error_MC_GCMM}, \ref{fig:reserve_rel_error_MC_pGCMM_b_t3} show Monte Carlo distributions of the relative reserve errors for LoB 1, LoB 2, and the total reserve under Settings 1, 2, and 3, respectively, based on 100 simulation runs. The reserve error distributions are generally symmetric and centered near zero across all three settings. For Settings 1 and 2, the relative reserve errors for LoB 1, LoB 2, and the total reserve exhibit similar distributional shapes with medians slightly above zero. Under the heavy-tailed random-effect setting (Setting 3), the reserve error distribution for LoB 2 displays mild right skewness, whereas the distributions for LoB 1 and the total reserve remain approximately symmetric. This suggests that heavy-tailed company-level heterogeneity has a greater impact on the more volatile LoB 2 than on LoB 1. However, the medians of the relative reserve errors remain close to zero across all three settings, indicating that the proposed estimator yields stable reserve performance despite changes in penalization and in the Gaussian random-effect assumption in the data-generating mechanisms.

We report TVaR at the 60\% level for the simulation study. For each setting, the Monte Carlo TVaR is computed from the empirical distribution of the actual total reserves obtained across the 100 simulation runs. For each simulation run, the actual reserve is computed by first converting the simulated lower-triangle loss ratios into losses using the corresponding earned premiums for each accident year. These losses are then aggregated across development years, accident years, companies, and the two LoBs to obtain the actual total reserve. The resulting 100 simulated total reserves are used to estimate the Monte Carlo TVaR$_{60}$. Table~\ref{tab:tvar_simulation} reports the relative error of the Monte Carlo TVaR$_{60}$ estimates under the three simulation settings. The relative errors are similar under Settings 1 and 2. Though Setting 3 exhibits a larger relative error for LoB 2 due to the heavy-tailed random effects, the relative error for the aggregate reserve remains close to 5\%, indicating that the proposed estimator provides reasonably stable tail-risk estimates under random-effect misspecification. 

The simulation study focuses on reserve accuracy, coefficient recovery, dependence estimation, and TVaR accuracy at a moderate tail level. Extending the simulation comparison to more extreme TVaR levels would require a substantially larger Monte Carlo design and is left for future robustness analysis.

\begin{table}[ht]
\centering
\caption{Relative error of TVaR estimates is computed as $(\widehat{\mathrm{TVaR}}_{60}-\mathrm{TVaR}_{60})/\mathrm{TVaR}_{60}$.}
\label{tab:tvar_simulation}
\begin{tabular}{lrrr}
\hline
Setting & LoB 1 & LoB 2 & Total \\
\hline
Setting 1 (pGCMM) & -6.01\% & -7.31\% & -6.10\% \\
Setting 2 (GCMM)  & -6.00\% & -7.37\% & -6.09\% \\
Setting 3 (Heavy-tailed RE) & -4.79\% & -22.22\% & -5.03\% \\
\hline
\end{tabular}
\end{table}




\color{black}
\newpage

\section{Summary and Discussion}

This paper develops a penalized copula mixed modeling framework for intercompany multivariate loss reserving and risk capital assessment. The proposed framework combines fixed effects for accident-year and development-year effects, company-specific random effects to capture insurer heterogeneity, and company-specific copula parameters for residual cross-LoB dependence. The penalized version further applies an $L_1$ penalty to weakly identified fixed effects, providing a more parsimonious and stable specification in the tail of the loss triangles.

The application highlights the importance of company-specific dependence in intercompany reserving. Appendix Tables~\ref{app:one_theta} and~\ref{app:one_theta_comparison} report the reserve point estimates and percentage errors under common-dependence specifications. These results show that allowing the copula parameter to vary by company substantially increases the flexibility of the copula benchmark. In the main application, the fixed-effect copula benchmark provides the closest total point reserve estimate for Company~1. However, the mixed-effect copula models provide more stable predictive distributions, lower coefficients of variation, and lower incremental tail capital. Taken together, these results suggest that the proposed pGCMM provides a credibility-type extension of copula regression that balances reserve prediction, intercompany borrowing strength, uncertainty quantification, and capital assessment.

The simulation study provides additional support for the proposed framework. Across controlled data-generating settings, the proposed framework recovers key model parameters, illustrates the effect of penalization, and remains informative under random-effect misspecification.  These findings complement the application results by demonstrating that the framework can produce stable predictive distributions and reliable tail-risk measures in addition to competitive reserve estimates.

Several limitations suggest directions for future work. First, the application and simulation study focus on two automobile lines of business. In higher-dimensional applications, the dependence component can be specified through an elliptical copula or a vine copula on the $\ell$-dimensional residual pseudo-observations. Estimation remains feasible via the same two-stage logic (fit marginals, then estimate dependence on residual ranks), with computational costs driven primarily by the copula dimension; in higher dimensions, structured dependence specifications, such as vine truncation or factor-copula approximations, can be adopted to preserve tractability. Second, point reserve estimates may remain sensitive to marginal specification and to the use of fixed accident-year and development-year factors, which do not fully capture the sequential nature of claims development. This motivates hybrid extensions that integrate recurrent architectures, such as gated recurrent units (GRUs), with copula-based dependence modeling. Such models can learn development dynamics while retaining interpretable dependence and capital assessment. Finally, future work could consider hierarchical structures for the copula parameters themselves, allowing dependence to vary across companies while still borrowing information across the market.

Another extension would be to relax the independence assumption for the company-specific random effects across LoBs and introduce a bivariate random-effect structure, allowing correlation between the random effects of different LoBs. In fact, in intercompany applications, insurers may also be exposed to common systemic factors (e.g., inflation, legal environment, or macroeconomic conditions), inducing dependence beyond the cell-level cross-LoB dependence captured by the copula. Our baseline formulation treats company effects as independent across insurers; extending the framework to incorporate correlated random effects, shared latent factors or market-level random effects  would provide a richer representation of dependence while preserving interpretability.

In summary, the proposed penalized generalized copula mixed model (pGCMM) provides a structured framework for intercompany multivariate reserving, with competitive reserve estimates and improved stability in predictive uncertainty and risk-capital assessment when multiple related triangles are available. Such data arise naturally in regulatory, supervisory, and rating contexts, as well as within large insurance groups across subsidiaries, segments, or regions. When only a single insurer is available, the framework reduces to a structured multivariate reserving model with the same dependence interpretation but without cross-entity borrowing of strength.

\section*{Acknowledgements}
This work was funded by the Start-up funds, the Faculty of Science at McMaster University (Pratheepa Jeganathan), the Natural Sciences and Engineering Research Council of Canada (NSERC), Discovery Grant [RGPIN-2022-05272] (Pratheepa Jeganathan), and Natural Sciences and Engineering Research Council of Canada (NSERC), Discovery Grant [No. 20016011] (Anas Abdallah). Support provided by SHARCNET (sharcnet.ca) and the Digital Research Alliance of Canada (alliancecan.ca) partly enabled this research.

\section*{Authors' Contribution}

Conceptualization, A.A. and P.J.; methodology, P.J., A.A., and P.C.; software, P.J. and P.C.; validation, P.J., A.A., and P.C.; formal analysis, P.J. and P.C.; investigation, A.A., P.J., and P.C.; resources, A.A. and P.J.; writing—original draft preparation,  A.A., P.J., and P.C.; writing—review and editing, A.A. and P.J.; supervision, A.A. and P.J.

\section*{Supplementary materials}
All R code related to the application and simulation results is maintained in a GitHub repository, which will be made publicly available upon acceptance of the manuscript. Prior to acceptance, the code is available from the corresponding author upon reasonable request.

\bibliography{manuscript2}

@article{Shi2011,
  title     = {Dependent Loss Reserving Using Copulas},
  author    = {Shi, Peng and Frees, Edward W.},
  journal   = {ASTIN Bulletin: The Journal of the IAA},
  volume    = {41},
  number    = {2},
  pages     = {449--486},
  year      = {2011}
}

@article{Kuo2019,
   title={DeepTriangle: A Deep Learning Approach to Loss Reserving},
   volume={7},
   ISSN={2227-9091},
   url={http://dx.doi.org/10.3390/risks7030097},
   DOI={10.3390/risks7030097},
   number={3},
   journal={Risks},
   publisher={MDPI AG},
   author={Kuo, Kevin},
   year={2019},
   month={Sep},
   pages={97}
}

@Article{Abdallah2015,
  author={Abdallah, Anas and Boucher, Jean-Philippe and Cossette, Helene},
  title={{Modeling dependence between loss triangles with Hierarchical Archimedean copulas}},
  journal={ASTIN Bulletin: The Journal of the IAA},
  year=2015,
  volume={45},
  number={3},
  pages={577-599},
  month={September}
}

@article{Abdallah2016,
author = {Abdallah, Anas and Boucher, Jean-Philippe and Cossette, Helene and Trufin, Julien},
year = {2016},
month = {04},
pages = {1-17},
title = {Sarmanov Family of Bivariate Distributions for Multivariate Loss Reserving Analysis},
volume = {20},
journal = {North American Actuarial Journal},
doi = {10.1080/10920277.2016.1161525}
}

@book{Nelsen_copula_2006,
author = {Nelsen, Roger B.},
title = {An Introduction to Copulas (Springer Series in Statistics)},
year = {2006},
isbn = {0387286594},
publisher = {Springer-Verlag},
address = {Berlin, Heidelberg}
}

@article{tibshirani1996regression,
  title={Regression shrinkage and selection via the lasso},
  author={Tibshirani, Robert},
  journal={Journal of the Royal Statistical Society: Series B (Methodological)},
  volume={58},
  number={1},
  pages={267--288},
  year={1996},
  publisher={Wiley Online Library}
}

@article{mcguire2018self,
  title={Self-assembling insurance claim models using regularized regression and machine learning},
  author={McGuire, Gr{\'a}inne and Taylor, Greg and Miller, Hugh},
  journal={Available at SSRN 3241906},
  year={2018}
}

@article{taylor2007synchronous,
  title={A synchronous bootstrap to account for dependencies between lines of business in the estimation of loss reserve prediction error},
  author={Taylor, Greg and McGuire, Gr{\'a}inne},
  journal={North American Actuarial Journal},
  volume={11},
  number={3},
  pages={70--88},
  year={2007},
  publisher={Taylor \& Francis}
}

@article{meyers2011loss,
  title={{Loss reserving data pulled from NAIC Schedule P}},
  author={Meyers, Glenn G and Shi, Peng},
  journal={URL: https://www.casact.org/publications-research/research/research-resources/loss-reserving-data-pulled-naic-schedule-p},
  volume={5},
  year={2011}
}

@article{abdallah2023rank,
  title={Rank-Based Multivariate Sarmanov for Modeling Dependence between Loss Reserves},
  author={Abdallah, Anas and Wang, Lan},
  journal={Risks},
  volume={11},
  number={11},
  pages={187},
  year={2023},
  publisher={MDPI}
}

@article{zhang2010general,
  title={A general multivariate chain ladder model},
  author={Zhang, Yanwei},
  journal={Insurance: Mathematics and Economics},
  volume={46},
  number={3},
  pages={588--599},
  year={2010},
  publisher={Elsevier}
}

@article{kirschner2008two,
  author = {Kirschner, G. and Kerley, C. and Isaacs, B.},
  title = {Two approaches to calculating correlated reserve indications across multiple lines of business},
  journal = {Variance},
  year = {2008},
  volume = {2},
  number = {1},
  pages = {15--38}
}

@book{davison1997bootstrap,
  title={Bootstrap methods and their application},
  author={Davison, Anthony Christopher and Hinkley, David Victor},
  number={1},
  year={1997},
  publisher={Cambridge university press}
}

@article{chatterjee2011bootstrapping,
  title={Bootstrapping lasso estimators},
  author={Chatterjee, Arindam and Lahiri, Soumendra Nath},
  journal={Journal of the American Statistical Association},
  volume={106},
  number={494},
  pages={608--625},
  year={2011},
  publisher={Taylor \& Francis}
}

@article{cai2025edt,
  title={Recurrent Neural Networks for Multivariate Loss Reserving and Risk Capital Analysis},
  author={Cai, Pengfei and Abdallah, Anas and Jeganathan, Pratheepa},
  journal={North American Actuarial Journal},
  pages={1--25},
  year={2025},
  publisher={Taylor \& Francis},
  doi = {10.1080/10920277.2025.2517149}
}

@article{karatacs2025dependence,
  title={Dependence modelling on loss triangles: copula regression with unobserved effect},
  author={Karata{\c{s}}, R{\"u}meysa and Karabey, U{\u{g}}ur},
  journal={Communications in Statistics-Simulation and Computation},
  pages={1--17},
  year={2025},
  publisher={Taylor \& Francis}
}

@article{shi2017multivariate,
  title={A multivariate analysis of intercompany loss triangles},
  author={Shi, Peng},
  journal={Journal of Risk and Insurance},
  volume={84},
  number={2},
  pages={717--737},
  year={2017},
  publisher={Wiley Online Library}
}

@article{antonio2010multilevel,
  title={A multilevel analysis of intercompany claim counts},
  author={Antonio, Katrien and Frees, Edward W and Valdez, Emiliano A},
  journal={ASTIN Bulletin: The Journal of the IAA},
  volume={40},
  number={1},
  pages={151--177},
  year={2010},
  publisher={Cambridge University Press}
}
\newpage
\begin{appendices}

\renewcommand{\thefigure}{\thesection.\arabic{figure}}
\setcounter{figure}{0}

\renewcommand{\thetable}{\thesection.\arabic{table}}
\setcounter{table}{0}

\renewcommand{\thealgorithm}{\thesection.\arabic{algorithm}}
\setcounter{algorithm}{0}

\section{Algorithms}

\subsection{Estimation Algorithm for GCMM}

\begin{algorithm}[H]
\caption{Iterative estimation procedure for the GCMM}
\label{alg:gcmm-estimation}
\begin{algorithmic}[1]
\State Input: Observed loss ratios
$\{Y_{ijc}^{(\ell)}\}$, design matrices $\{\boldsymbol{x}_{ij}^{(\ell)}\}$,
number of Monte Carlo samples $M$, and tolerances $\varepsilon_1, \varepsilon_2$.

\State Initialization: Fit separate marginal models for each LoB to obtain initial values for
$\boldsymbol{\beta}^{(1)},\boldsymbol{\beta}^{(2)},\sigma_1,\sigma_2,\tau_1,\tau_2$. Set $\theta_c^{(0)}=0$ for all $c=1,\ldots,C$.

\For{$r=0,1,2,\ldots$}

    \State \parbox[t]{0.9\linewidth}{ Step 1: Update marginal mixed-model parameters. Given the current copula parameters $\boldsymbol{\theta}^{(r)}=(\theta_1^{(r)},\ldots,\theta_C^{(r)})$, draw Monte Carlo samples $b_{c,m}^{(\ell)} \sim N(0,\tau_\ell^2)$, $m=1,\ldots,M$, and update $$\boldsymbol{\Theta}_{-\theta_c}^{(r+1)}=\arg\max_{\boldsymbol{\Theta}_{-\theta_c}}\mathcal{L}(\boldsymbol{\Theta}_{-\theta_c}),$$ where $\mathcal{L}(\boldsymbol{\Theta}_{-\theta_c})$ is approximated as in Eq.~\eqref{eq:app-step1}.}

    \State \parbox[t]{0.9\linewidth}{Step 2: Update company-specific copula parameters.}

    \State \hspace{\algorithmicindent}  \parbox[t]{0.85\linewidth}{
    Using the updated marginal parameter estimates $\boldsymbol{\Theta}_{-\theta_c}^{(r+1)}$, compute pseudo-residuals
    $\epsilon_{ijc}^{(\ell)}$ for each LoB. For a log-normal margin, use
    Eq.~\eqref{eqn:lognormal_residual}; for a gamma margin, use
    Eq.~\eqref{eqn:gamma_residual}.
    }

     \State \hspace{\algorithmicindent}  
     \parbox[t]{0.85\linewidth}{ Transform the pseudo-residuals into rank-based pseudo-observations using Eq.~\eqref{eq:pseudo_res}.}

     \State \hspace{\algorithmicindent} 
     \parbox[t]{0.85\linewidth}{ With the marginal model parameters fixed, update each company-specific copula parameter by maximizing the rank-based pseudo-log-likelihood: $$\boldsymbol{\theta}^{(r+1)}=\arg\max_{\boldsymbol{\theta}}
    \mathcal{L}(\boldsymbol{\theta}),$$ where $\mathcal{L}(\boldsymbol{\theta})$ is given in Eq.~\eqref{eq:appr-Step2}.}

    \State Convergence check: Stop if Eq.~\eqref{eq:conv_pars} and Eq.~\eqref{eq:conv_theta} satisfied.

\EndFor

\State Output:
$\widehat{\boldsymbol{\Theta}}_{-\theta_c} =\lbrace \hat{\boldsymbol{\beta}}^{(1)},\hat{\boldsymbol{\beta}}^{(2)},\hat{\sigma}_1,\hat{\sigma}_2,\hat{\tau}_1,\hat{\tau}_2 \rbrace$ and
$\widehat{\boldsymbol{\theta}}=(\hat{\theta}_1,\ldots,\hat{\theta}_C)$.
\end{algorithmic}
\end{algorithm}

\subsection{Estimation Algorithm for pGCMM}

\begin{algorithm}[H]
\caption{Iterative estimation procedure for the penalized GCMM}
\label{alg:pgcmm-estimation}
\begin{algorithmic}[1]
\State Input: Observed loss ratios
$\{Y_{ijc}^{(\ell)}\}$, design matrices $\{\boldsymbol{x}_{ij}^{(\ell)}\}$,
number of Monte Carlo samples $M$, penalty parameters $\lambda_1,\lambda_2$,
and tolerances $\varepsilon_1, \varepsilon_2$.

\State Initialization: Fit separate marginal models for each LoB to obtain initial values for
$\boldsymbol{\beta}^{(1)},\boldsymbol{\beta}^{(2)},\sigma_1,\sigma_2,\tau_1,\tau_2$. Set $\theta_c^{(0)}=0$ for all $c=1,\ldots,C$.

\For{$r=0,1,2,\ldots$}

    \State \parbox[t]{0.9\linewidth}{ Step 1: Update penalized marginal mixed-model parameters. Given the current copula parameters $\boldsymbol{\theta}^{(r)}=(\theta_1^{(r)},\ldots,\theta_C^{(r)})$, draw Monte Carlo samples $b_{c,m}^{(\ell)} \sim N(0,\tau_\ell^2)$, $m=1,\ldots,M$, and update $$\boldsymbol{\Theta}_{-\theta_c}^{(r+1)}=\arg\min_{\boldsymbol{\Theta}_{-\theta_c}}J_{\lambda_1,\lambda_2}(\boldsymbol{\Theta}_{-\theta_c}),$$ where $J_{\lambda_1,\lambda_2}(\boldsymbol{\Theta}_{-\theta_c})$ is approximated as in Eq.~\eqref{eq:app-pen-neg-loglik}.}

    \State \parbox[t]{0.9\linewidth}{Step 2: Update company-specific copula parameters.}

    \State \hspace{\algorithmicindent}  \parbox[t]{0.85\linewidth}{
    Using the updated penalized marginal parameter estimates $\boldsymbol{\Theta}_{-\theta_c}^{(r+1)}$, compute pseudo-residuals
    $\epsilon_{ijc}^{(\ell)}$ for each LoB. For a log-normal margin, use
    Eq.~\eqref{eqn:lognormal_residual}; for a gamma margin, use
    Eq.~\eqref{eqn:gamma_residual}.
    }

     \State \hspace{\algorithmicindent}  
     \parbox[t]{0.85\linewidth}{ Transform the pseudo-residuals into rank-based pseudo-observations using Eq.~\eqref{eq:pseudo_res}.}

     \State \hspace{\algorithmicindent} 
     \parbox[t]{0.85\linewidth}{ With the marginal model parameters fixed, update each company-specific copula parameter by maximizing the rank-based pseudo-log-likelihood: $$\boldsymbol{\theta}^{(r+1)}=\arg\max_{\boldsymbol{\theta}}
    \mathcal{L}(\boldsymbol{\theta}),$$ where $\mathcal{L}(\boldsymbol{\theta})$ is given in Eq.~\eqref{eq:appr-Step2}.}

    \State Convergence check: Stop if Eq.~\eqref{eq:conv_pars} and Eq.~\eqref{eq:conv_theta} satisfied.

\EndFor

\State Output:
$\widehat{\boldsymbol{\Theta}}_{-\theta_c} =\lbrace \hat{\boldsymbol{\beta}}^{(1)},\hat{\boldsymbol{\beta}}^{(2)},\hat{\sigma}_1,\hat{\sigma}_2,\hat{\tau}_1,\hat{\tau}_2 \rbrace$ and
$\widehat{\boldsymbol{\theta}}=(\hat{\theta}_1,\ldots,\hat{\theta}_C)$.
\end{algorithmic}
\end{algorithm}

\subsection{Modified Bootstrap Procedure}

\begin{algorithm}[H]
\caption{Modified bootstrap procedure for the predictive distribution of the reserve}
\label{alg:pgcmm-bootstrap}
\begin{algorithmic}[1]

\State Input: Observed incremental paid losses $\{y_{ijc}^{(\ell)}\}$,
design matrices $\{\boldsymbol{x}_{ij}^{(\ell)}\}$, number of bootstrap samples $B$,
and number of Monte Carlo samples $M$.

\State Fit the pGCMM to the observed data and obtain
$\hat{\lambda}_1,\hat{\lambda}_2$,
$\hat{\boldsymbol{\beta}}^{(\ell)}$, $\hat{\sigma}_\ell$,
$\hat{\tau}_\ell$, and $\hat{\theta}_c$, $c=1,\ldots,C$.

\State Construct thresholded coefficient estimates
$\tilde{\boldsymbol{\beta}}^{(\ell)}$ by setting coefficients with absolute
values less than $0.001$ to zero.

\State Compute posterior mode estimates $\hat b_c^{(\ell)}$ of the company-specific
random effects and fitted means $$\tilde{\mu}_{ijc}^{(\ell)}=\exp\!\left(
\boldsymbol{x}_{ij}^{(\ell)}\tilde{\boldsymbol{\beta}}^{(\ell)}+\hat{b}_c^{(\ell)}\right).$$

\State Compute pseudo-residuals using the marginal-model-specific residual
definition in Eq.~\eqref{eqn:lognormal_residual} or Eq.~\eqref{eqn:gamma_residual}.

\For{$b=1,\ldots,B$}

    \For{$c=1,\ldots,C$}

        \State Simulate dependent pairs
        $(u_{ijc}^{*(1)},u_{ijc}^{*(2)})$ from the fitted copula
        $C(\cdot;\hat{\theta}_c)$.

        \State \parbox[t]{0.85\linewidth}{ Transform the simulated uniforms using the empirical quantile functions of the pseudo-residuals to obtain $\epsilon_{ijc}^{*(\ell)}$, $\ell=1,2$.}

        \State \parbox[t]{0.85\linewidth}{ Generate bootstrap incremental paid losses $y_{ijc}^{*(\ell)}$ by applying the inverse transformation associated with the selected marginal model. For example, under the gamma margin, $$y_{ijc}^{*(\ell)}=\epsilon_{ijc}^{*(\ell)}\left(\tilde{\mu}_{ijc}^{(\ell)}\hat{\sigma}_{\ell}\right)^{1/2} + \tilde{\mu}_{ijc}^{(\ell)}.$$}

    \EndFor

    \State \parbox[t]{0.85\linewidth}{ Refit the pGCMM to the bootstrap sample using the fixed penalization parameters $\hat{\lambda}_1$ and $\hat{\lambda}_2$.}

    \State \parbox[t]{0.85\linewidth}{ Predict the lower-triangle incremental losses and compute the bootstrap
    reserve for company $c$: $$R_c^{*(b)}=\sum_{\ell=1}^{2}\sum_{i=2}^{I}\sum_{j=I-i+2}^{I}\omega_{ic}^{(\ell)}\exp\!\left(\hat{\eta}_{ijc}^{*(\ell)}
    \right),$$ where $\hat{\eta}_{ijc}^{*(\ell)} = \boldsymbol{x}_{ij}^{(\ell)} \hat{\boldsymbol{\beta}}^{*(\ell)}+\hat b_c^{*(\ell)}$.}

\EndFor

\State Output: Bootstrap reserve replicates
$\{R_c^{*(b)}: b=1,\ldots,B\}$ for each company $c$.

\end{algorithmic}
\end{algorithm}

\renewcommand{\thefigure}{\thesection.\arabic{figure}}
\setcounter{figure}{0}

\renewcommand{\thetable}{\thesection.\arabic{table}}
\setcounter{table}{0}

\renewcommand{\thealgorithm}{\thesection.\arabic{algorithm}}
\setcounter{algorithm}{0}

\section{Application Results Contd.}\label{application_appendix}
For out-of-sample validation, we used a company-level late-development holdout scheme. Starting from the observed upper triangle, we first identified cells that were observed for both lines of business. We then randomly selected five companies and held out their last five observed development lags. The remaining observed cells were used for model fitting. For pGCMM, the regularization parameter $\lambda$ was chosen by minimizing the AIC on the training set. The resulting pGCMM and GCMM were then compared on the validation set, consisting of the last five observed development lags from five randomly selected companies, using RMSE and MAE as measures of predictive accuracy as in Table \ref{tab:validation}. The validation results indicate that pGCMM and GCMM achieve nearly identical predictive performance on the held-out development lags, but pGCMM yields a parsimonious mean structure.  After this, we trained four different models on all the upper triangle data and interpret the results.

\begin{table}[H]
\centering
\caption{Validation performance on the held-out development lags.}
\label{tab:validation}
\begin{tabular}{lccccc}
\toprule
Model & RMSE (LoB 1) & MAE (LoB 1) & RMSE (LoB 2) & MAE (LoB 2) \\
\midrule
GCMM  & 0.00653 & 0.00461 & 0.01950 & 0.01340 \\
pGCMM & 0.00653 & 0.00461 & 0.01950 & 0.01340 \\
\bottomrule
\end{tabular}
\end{table}

The following figures present the corresponding results for the GCMM, facilitating comparison with those obtained under the pGCMM.

\begin{figure}[H]
    \centering
    \includegraphics[width=0.8\textwidth]{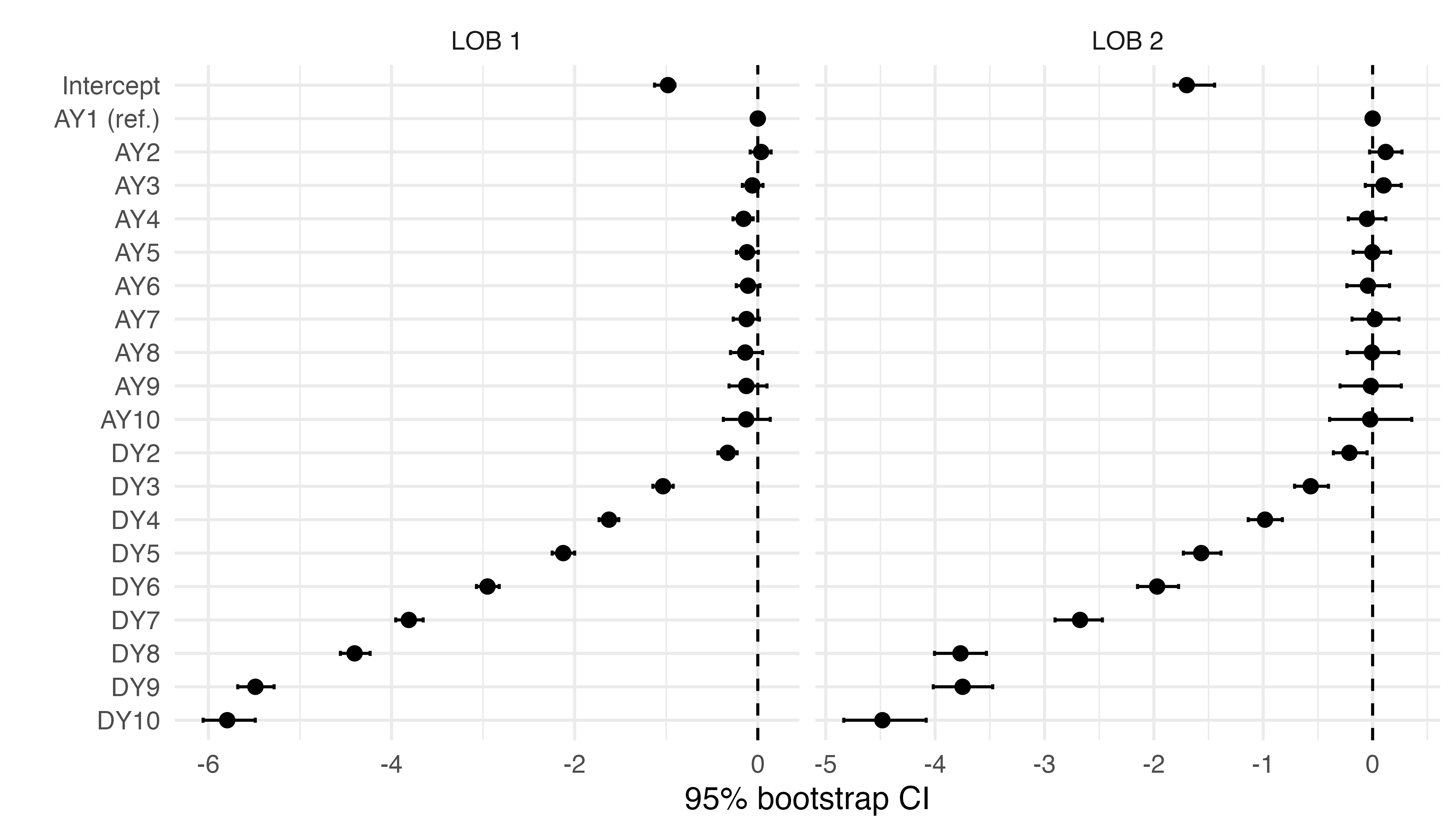}
    \caption{Point estimates and 95\% bootstrap confidence intervals for the fixed-effect parameters in LoB 1 and LoB 2 obtained from the fitted GCMM. }
    \label{fig:beta_bootstrap_CI_GCMM_nolasso}
\end{figure}

\begin{figure}[H]
    \centering
    \includegraphics[width=0.8\textwidth]{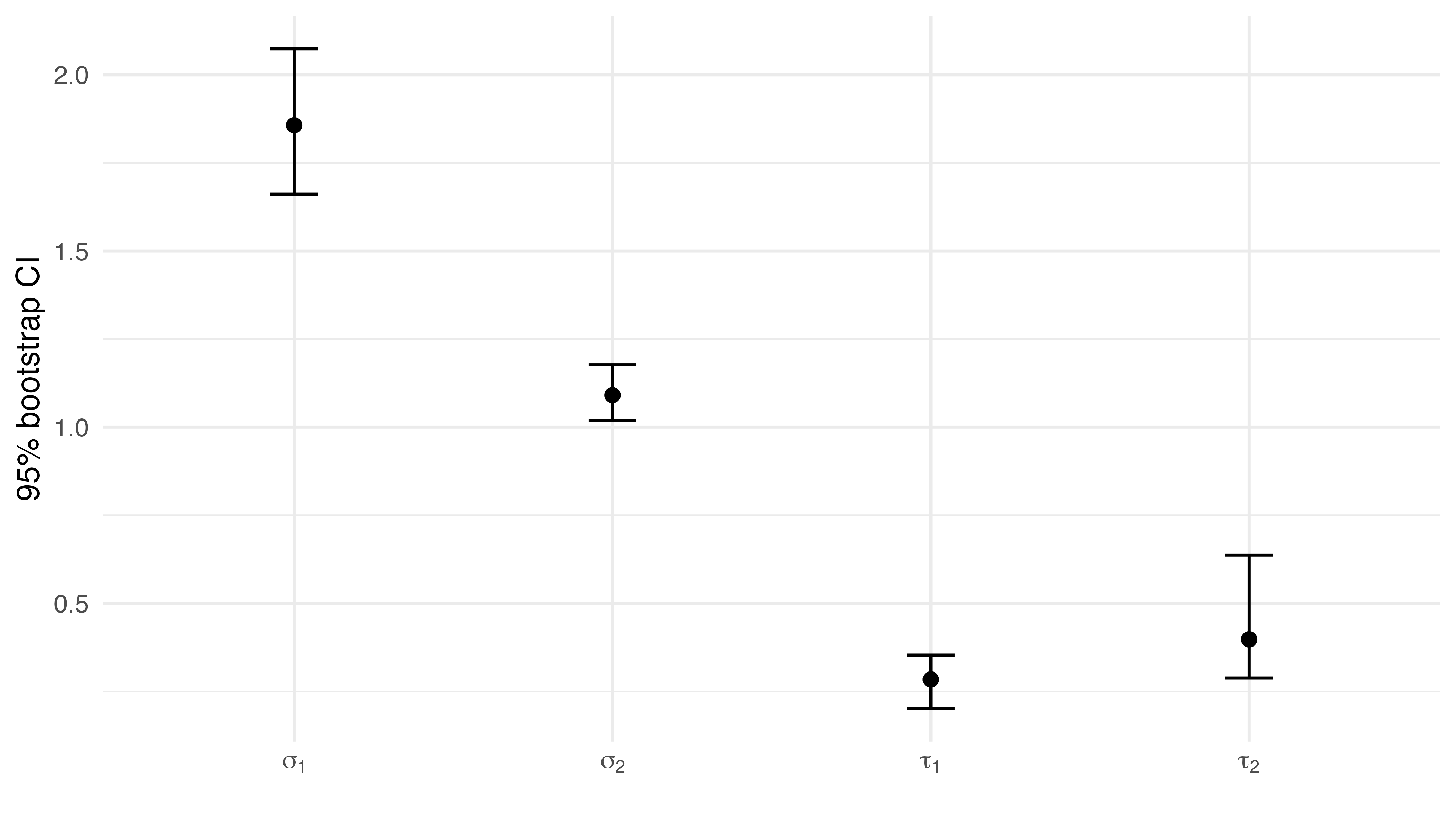}
    \caption{Point estimates and 95\% bootstrap confidence intervals for the gamma shape and random effect standard deviation for LoB 1 and LoB 2 obtained from the fitted GCMM.}
    \label{fig:variance_shape_bootstrap_CI_GCMM_nolasso}
\end{figure}

\begin{figure}[H]
    \centering
    \includegraphics[width=0.8\textwidth]{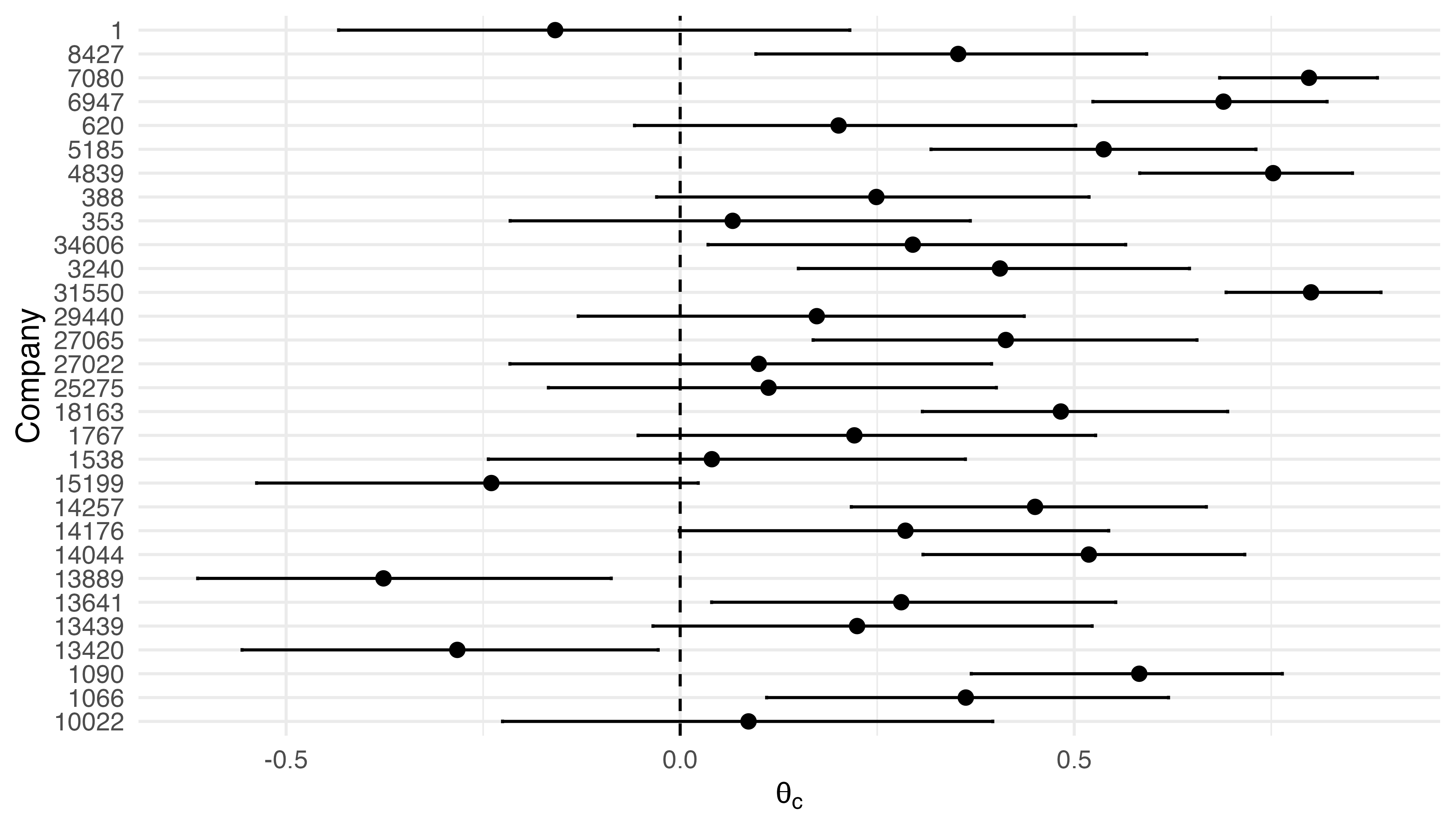}
    \caption{Point estimates and 95\% bootstrap confidence intervals for the company-specific copula dependence parameters $\theta_c$ obtained from the fitted GCMM. The dashed vertical line indicates zero dependence. Most companies exhibit positive dependence between the two lines of business, although the strength of dependence varies substantially across companies. For Company 1, the estimated dependence parameter is negative, consistent with previous analyses. However, the corresponding bootstrap confidence interval contains zero, indicating substantial estimation uncertainty and suggesting that the evidence for negative dependence is weak. This uncertainty may arise from the limited amount of company-specific information available for estimating dependence, particularly when the marginal variability in one or both lines of business is large compared to the LoB pairwise dependence.}
    \label{fig:theta_bootstrap_CI_GCMM_nolasso}
\end{figure}

\begin{figure}[H]
    \centering
    \includegraphics[width=0.8\textwidth]{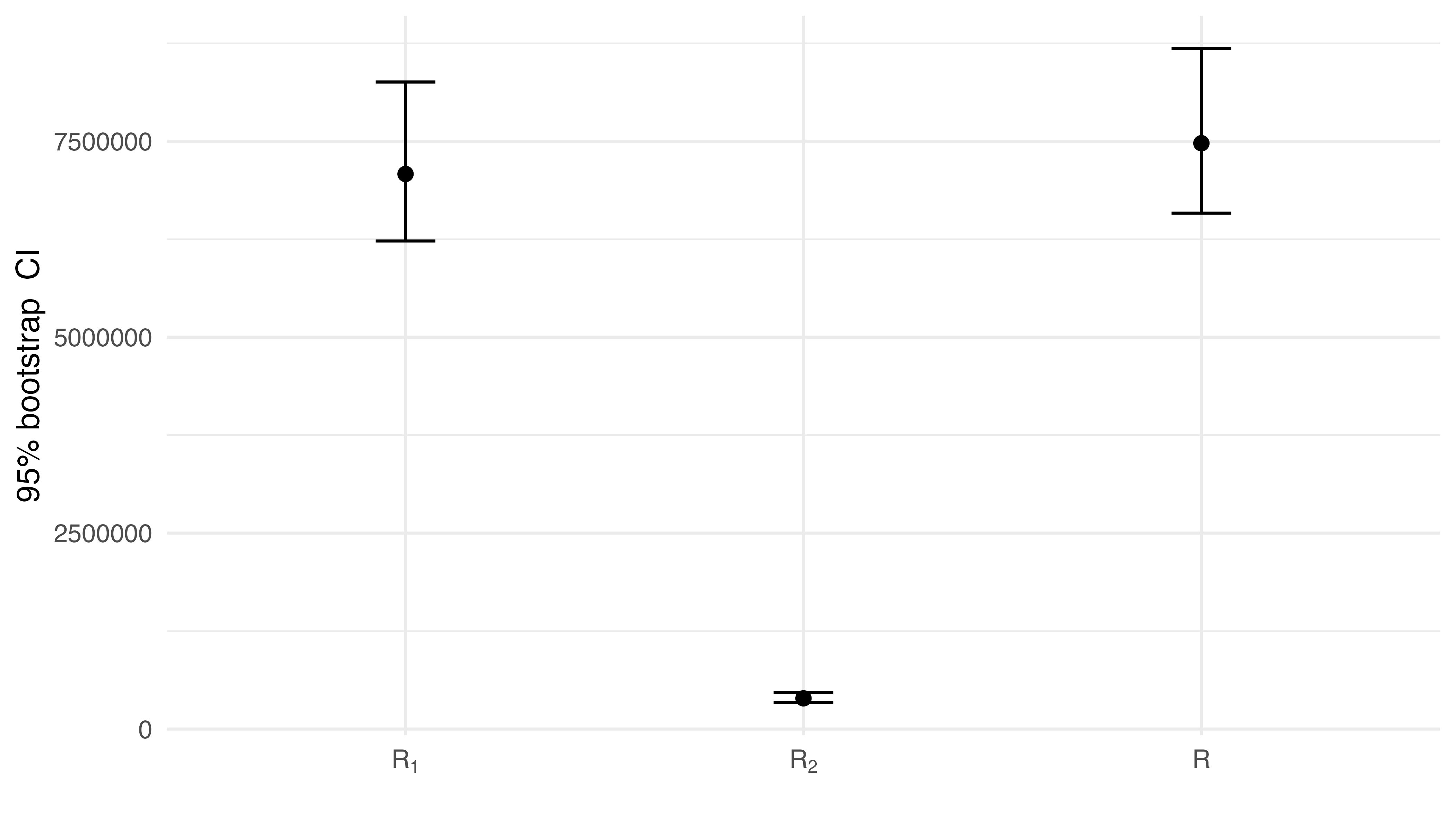}
    \caption{Point estimates and 95\% bootstrap confidence intervals for the reserves of LoB~1 ($R_1$), LoB~2 ($R_2$), and the total reserve ($R$) for Company~1 under the fitted GCMM.}
    \label{fig:reserve_bootstrap_CI_GCMM_Allasso}
\end{figure}


Table~\ref{app:one_theta} reports reserve point estimates for the corresponding common-dependence specifications, in which a single copula parameter is assumed for all companies.

\begin{table}[H]
\centering
\caption{Point estimates of unpaid loss reserves for Company 1 under the common $\theta$ for all companies assumption. Here, $R_1$ and $R_2$ denote reserves for LoB 1 and LoB 2, respectively, and $R=R_1+R_2$ denotes the total reserve.}
\label{app:one_theta}
\begin{tabular}[t]{l r r r}
\hline
Model & LoB 1, $R_1$ & LoB 2, $R_2$ & Total, $R$\\
\hline
 pGCMM ($\theta$) & 7,212,610 & 381,761 & 7,594,371\\

  GCMM ($\theta$) & 7,093,725 & 372,224 & 7,465,949\\

 Copula-FE ($\theta$) & 6,847,763 & 370,386 & 7,218,150\\

Actual reserve & 8,086,094 & 318,380 & 8,404,474\\
\hline
\end{tabular}
\end{table}

\begin{table}[H]
\centering
\caption{Percentage error of unpaid loss reserve estimates for Company 1 under the common $\theta$ for all companies assumption. Percentage error is computed as $(\widehat{R}-R)/R \times 100$, where $R$ denotes the actual reserve and $\widehat{R}$ the estimated reserve. Positive values indicate overestimation and negative values indicate underestimation.}
\label{app:one_theta_comparison}
\begin{tabular}[t]{ l r r r}
\hline
  Model & Personal Auto & Commercial Auto & Total\\
\hline
pGCMM ($\theta$) & -10.80\% & 19.91\% & -9.64\%\\

GCMM ($\theta$) & -12.27\% & 16.91\% & -11.17\%\\

 Copula-FE ($\theta$) & -15.31\% & 16.33\% & -14.12\%\\
\hline
\end{tabular}
\end{table}

\renewcommand{\thefigure}{\thesection.\arabic{figure}}
\setcounter{figure}{0}

\renewcommand{\thetable}{\thesection.\arabic{table}}
\setcounter{table}{0}

\renewcommand{\thealgorithm}{\thesection.\arabic{algorithm}}
\setcounter{algorithm}{0}

\section{Simulation Results}

\begin{figure}[H]
    \centering
    \includegraphics[width=0.8\textwidth]{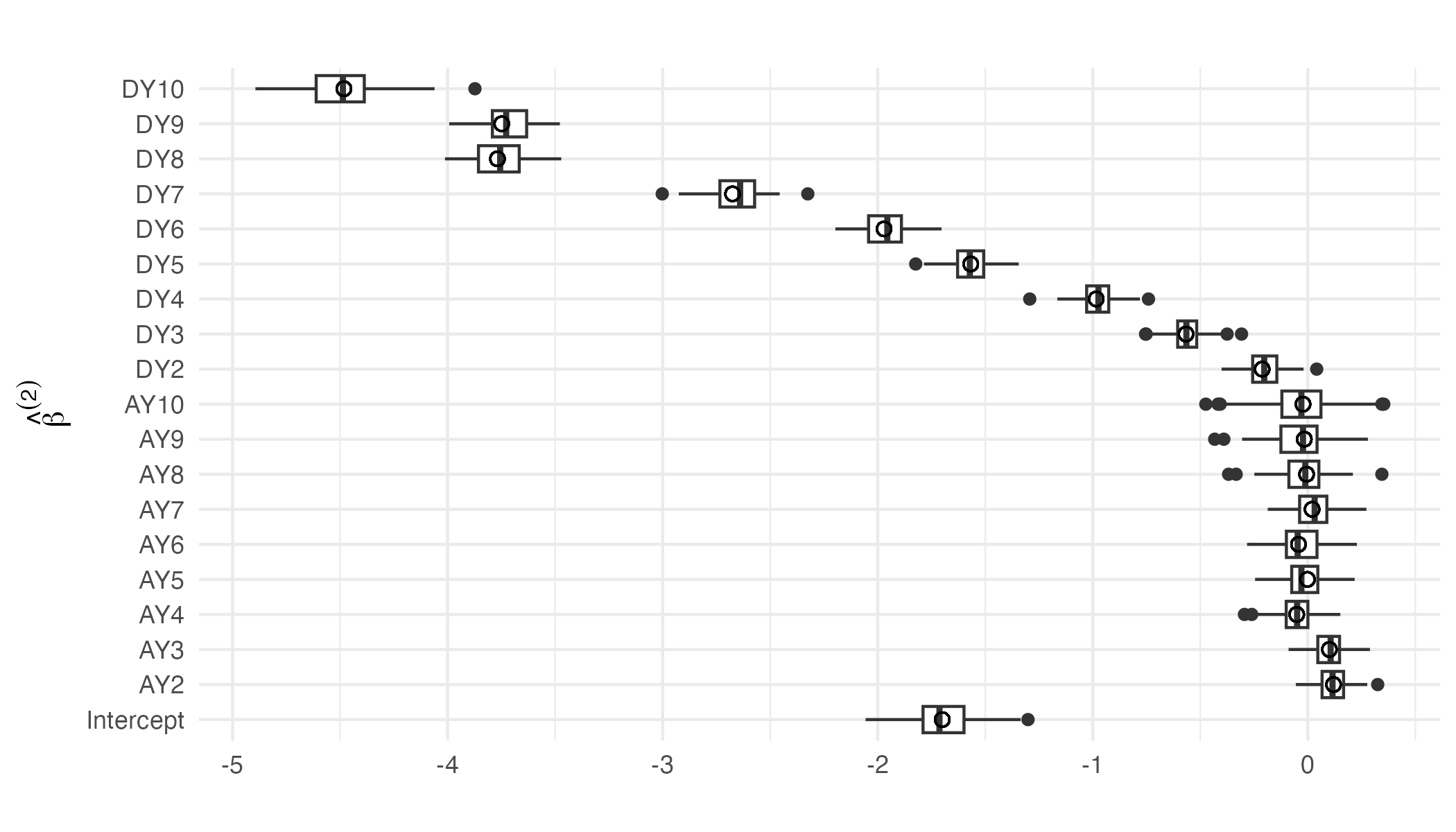}
    \caption{Monte Carlo sampling distributions of the estimated fixed effects for LoB 2 under Setting 2, based on 100 simulation runs. Boxplots summarize the parameter estimates, while open circles denote the actual values.}
    \label{fig:fixed_effects_LOB2_MC_GCMM}
\end{figure}

\begin{figure}[H]
    \centering
    \includegraphics[width=0.8\textwidth]{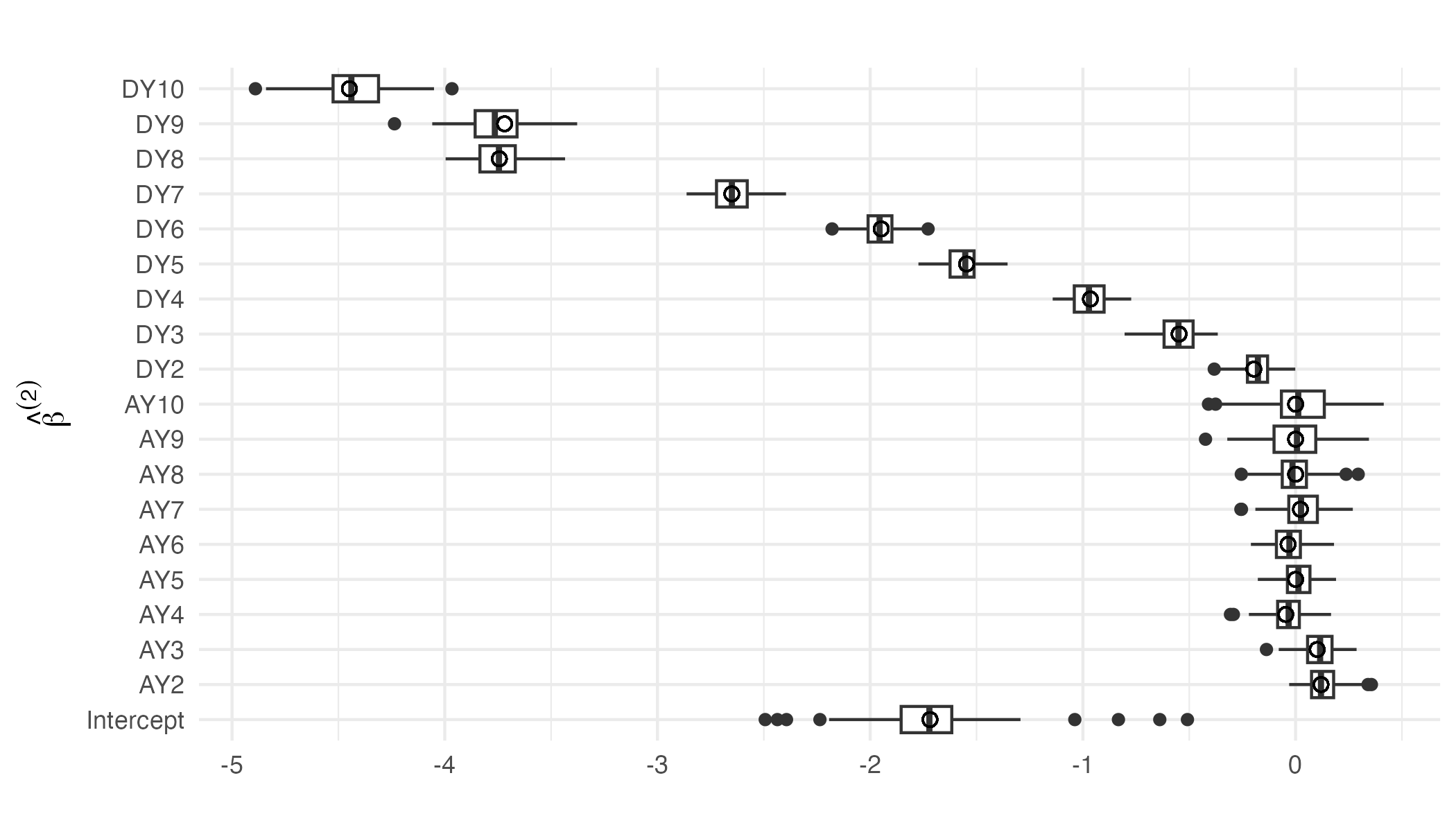}
    \caption{Monte Carlo sampling distributions of the estimated fixed effects for LoB 2 under Setting 3, based on 100 simulation runs. Boxplots summarize the parameter estimates, while open circles denote the actual values.}
    \label{fig:fixed_effects_LOB2_MC_pGCMM_b_t3}
\end{figure}

\begin{figure}[H]
    \centering
    \includegraphics[width=0.8\textwidth]{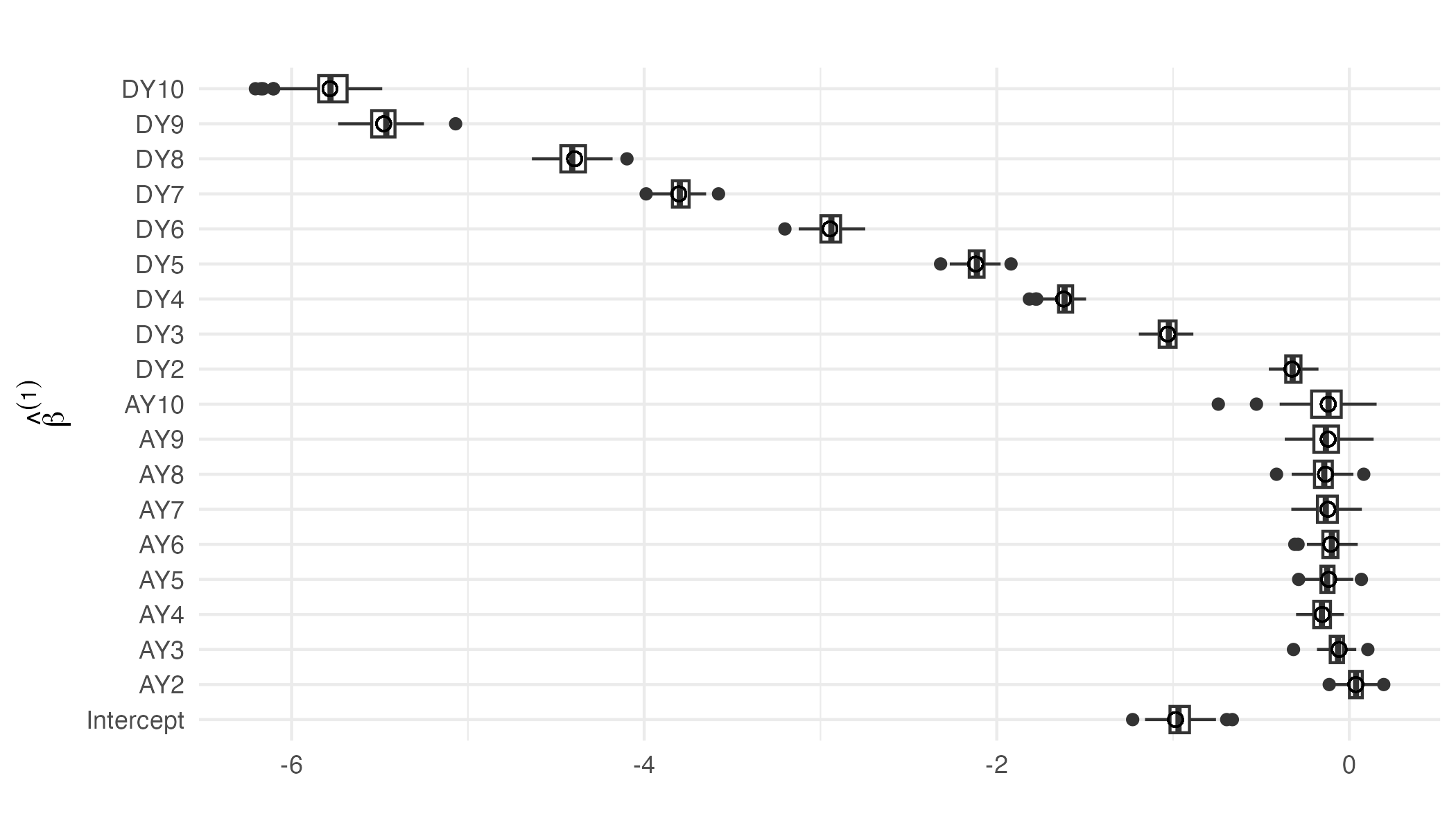}
    \caption{Monte Carlo sampling distributions of the estimated fixed effects for LoB 1 under Setting 1, based on 100 simulation runs. Boxplots summarize the parameter estimates, while open circles denote the actual values.}
    \label{fig:fixed_effects_LOB1_MC_pGCMM}
\end{figure}

\begin{figure}[H]
    \centering
    \includegraphics[width=0.8\textwidth]{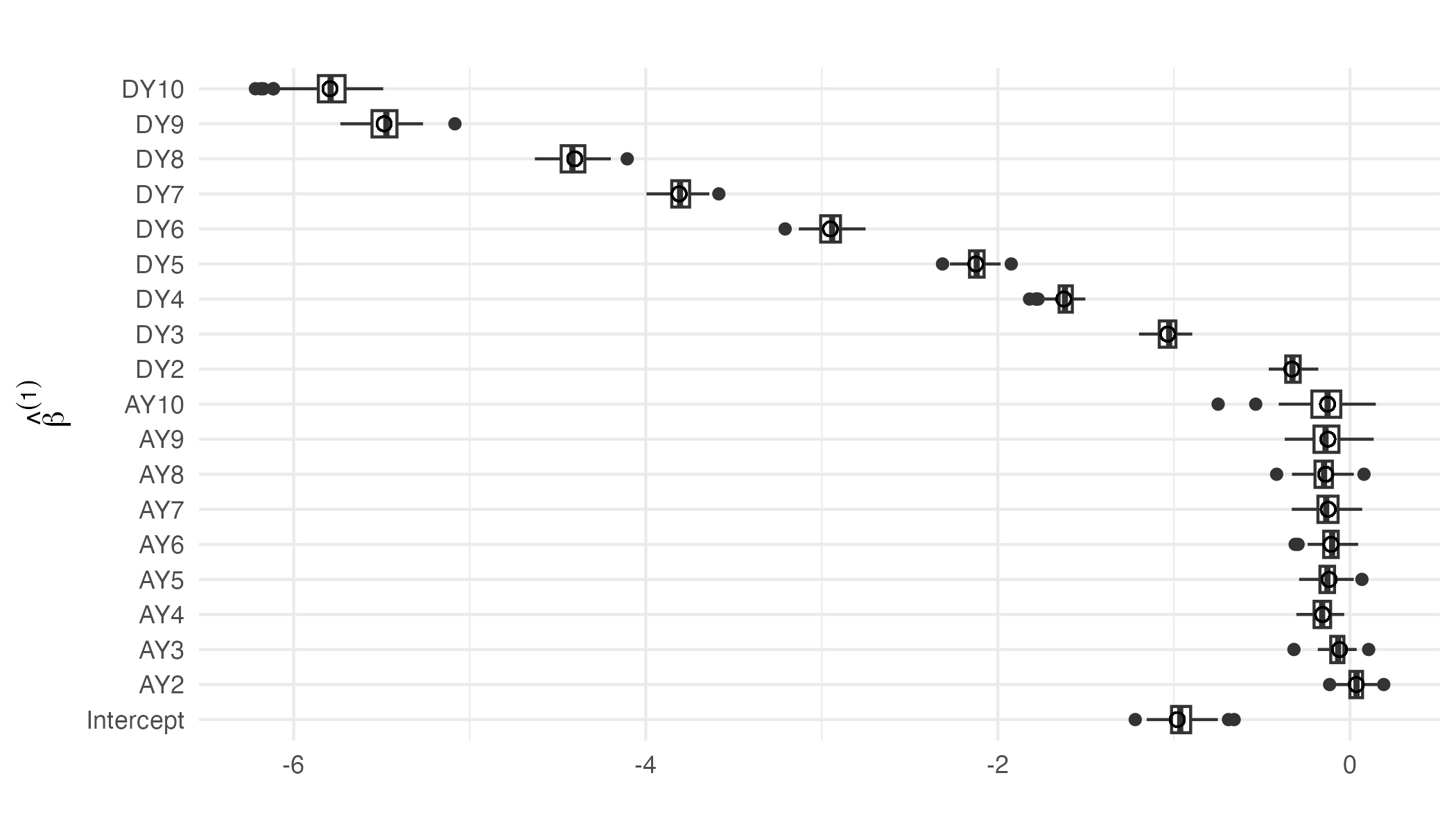}
    \caption{Monte Carlo sampling distributions of the estimated fixed effects for LoB 1 under Setting 2, based on 100 simulation runs. Boxplots summarize the parameter estimates, while open circles denote the actual values.}
    \label{fig:fixed_effects_LOB1_MC_GCMM}
\end{figure}

\begin{figure}[H]
    \centering
    \includegraphics[width=0.8\textwidth]{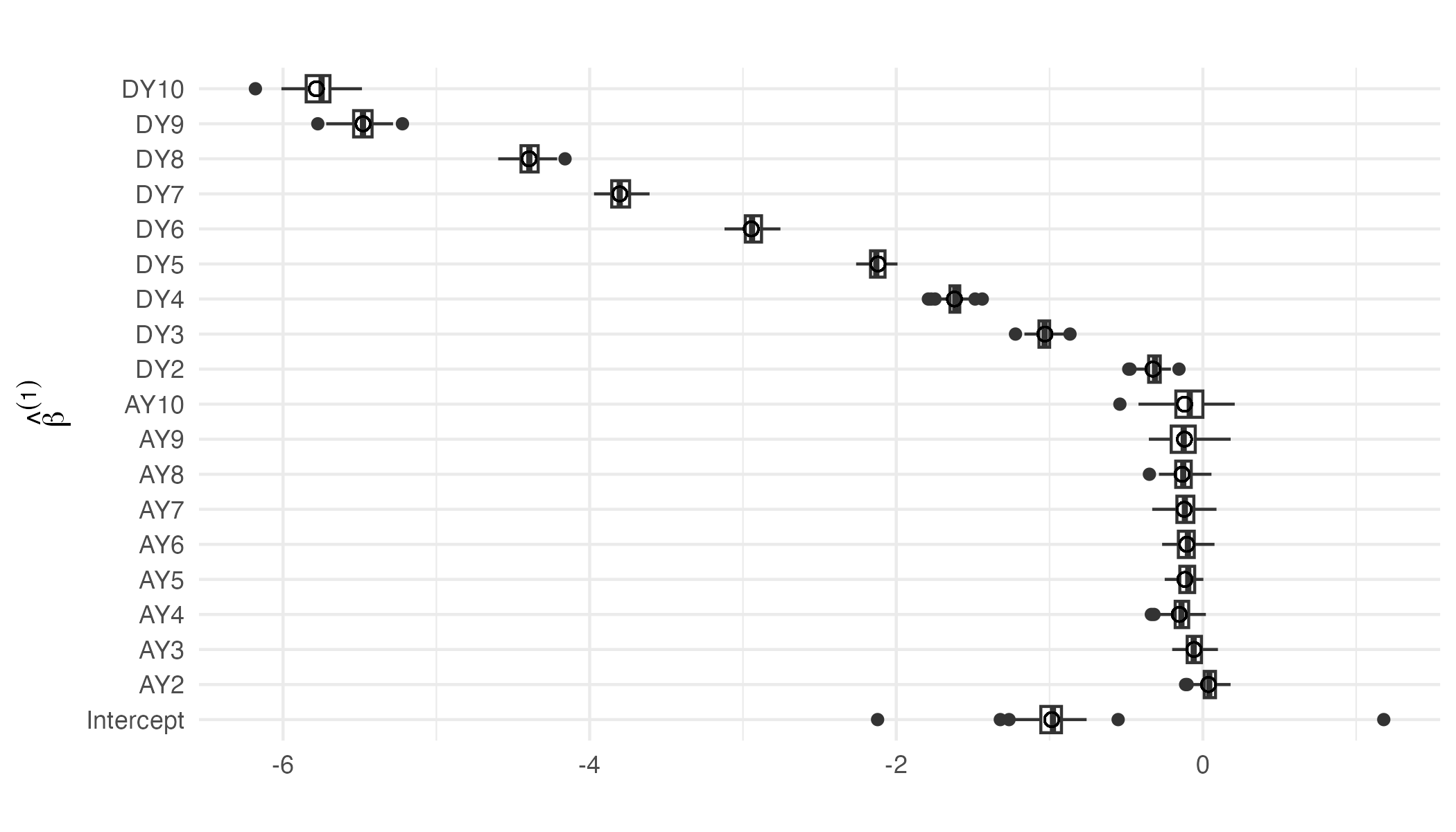}
    \caption{Monte Carlo sampling distributions of the estimated fixed effects for LoB 1 under Setting 3, based on 100 simulation runs. Boxplots summarize the parameter estimates, while open circles denote the actual values.}
    \label{fig:fixed_effects_LOB1_MC_pGCMM_b_t3}
\end{figure}

\begin{figure}[H]
    \centering
    \includegraphics[width=0.8\textwidth]{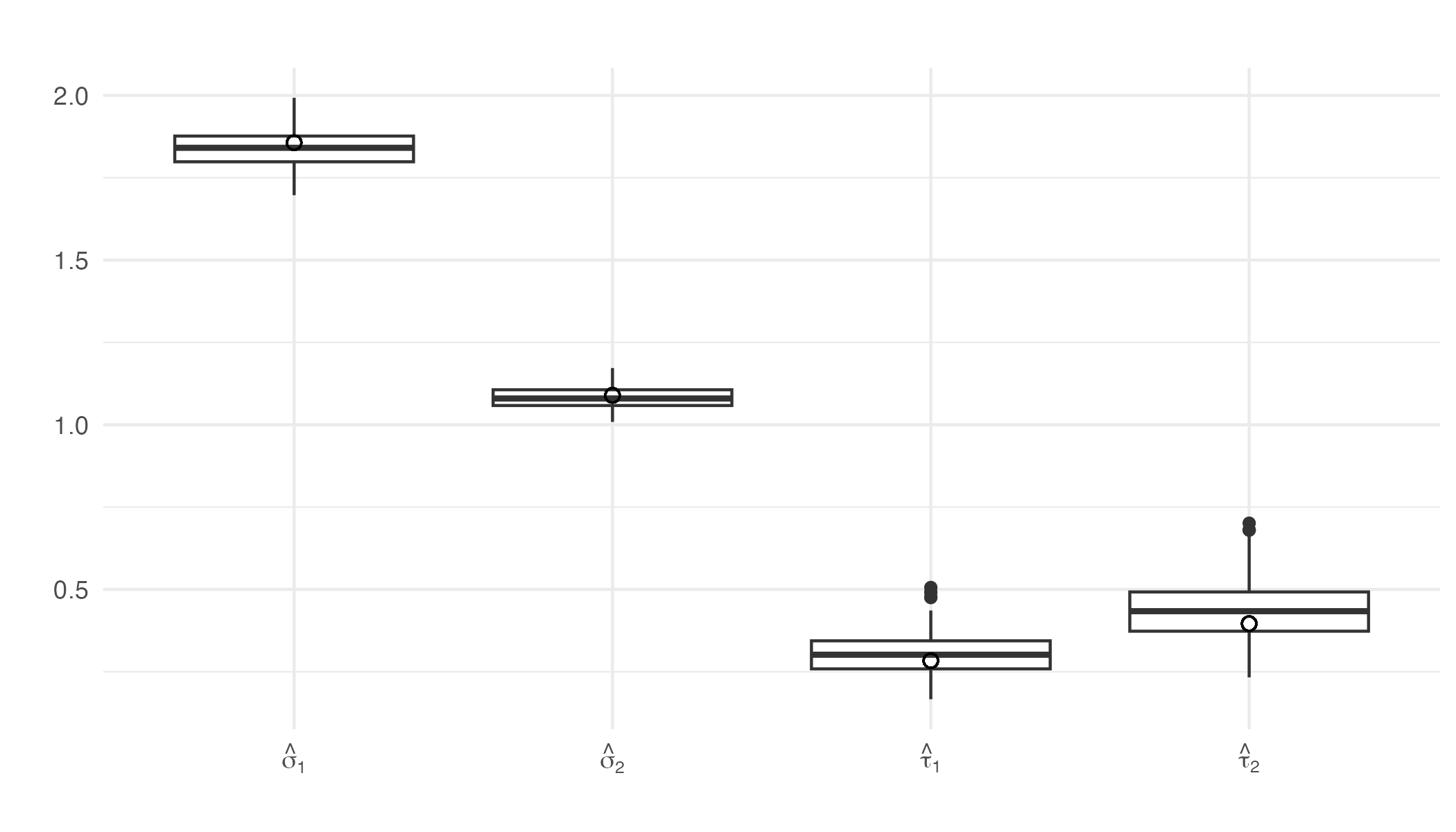}
    \caption{Monte Carlo sampling distributions of the estimated gamma shape parameters and random-effect standard deviations under Setting 1, based on 100 simulation runs. Boxplots summarize the parameter estimates, while open circles denote the actual values. }
    \label{fig:var_shape_MC_pGCMM}
\end{figure}

\begin{figure}[H]
    \centering
    \includegraphics[width=0.8\textwidth]{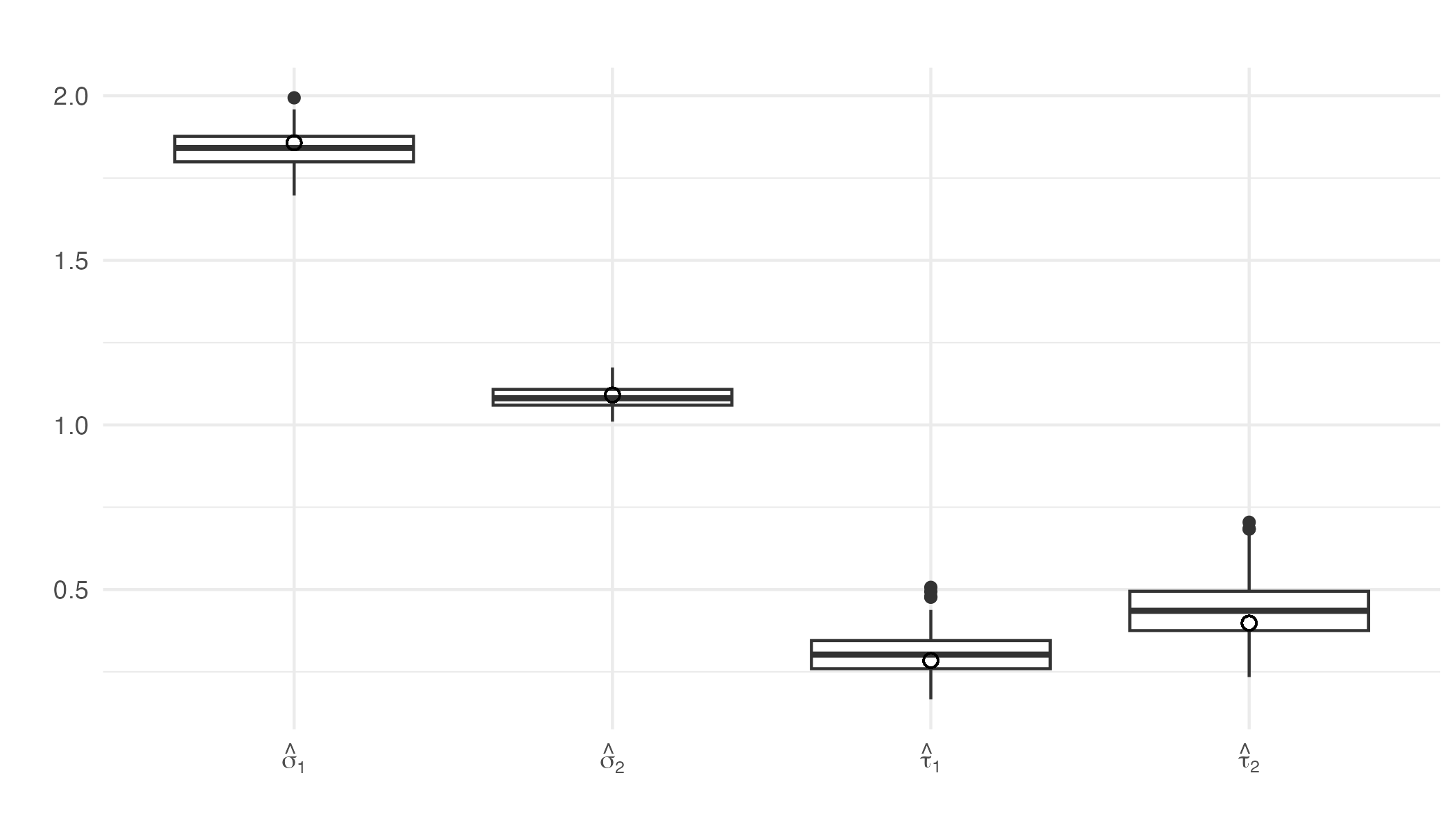}
    \caption{Monte Carlo sampling distributions of the estimated gamma shape parameters and random-effect standard deviations under Setting 2, based on 100 simulation runs. Boxplots summarize the variability of the estimates, while open circles denote the generating values. }
    \label{fig:var_shape_MC_GCMM}
\end{figure}

\begin{figure}[H]
    \centering
    \includegraphics[width=0.8\textwidth]{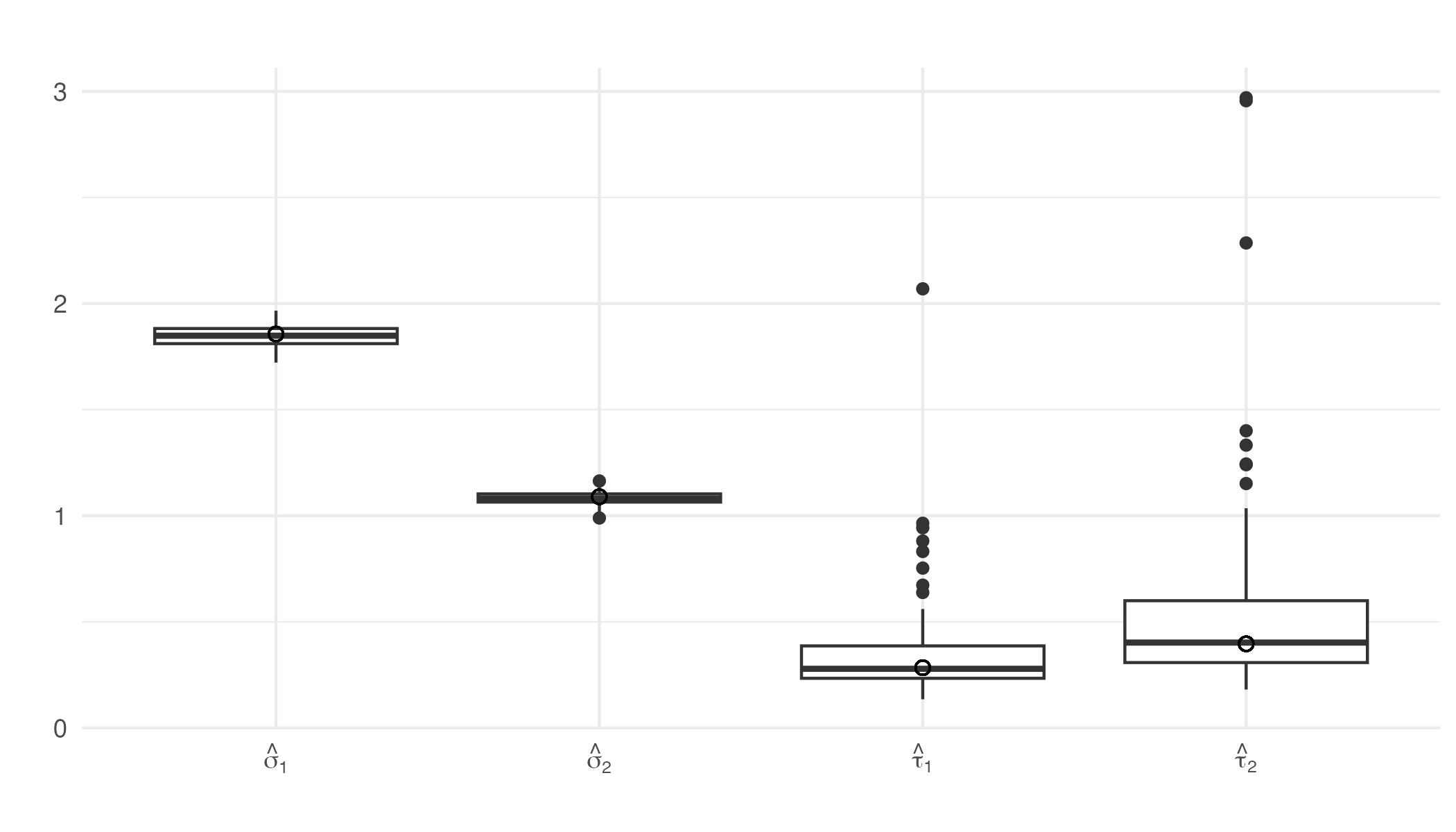}
    \caption{Monte Carlo sampling distributions of the estimated gamma shape parameters and random-effect standard deviations under Setting 3, based on 100 simulation runs. Boxplots summarize the variability of the estimates, while open circles denote the generating values. }
    \label{fig:var_shape_MC_pGCMM_b_t3}
\end{figure}

\end{appendices}

\end{document}